\documentclass[fleqn,usenatbib,onecolumn]{mnras} 
\usepackage{newtxtext,newtxmath}
\usepackage[T1]{fontenc}
\usepackage{ae,aecompl}

\usepackage{mathtools} 
\usepackage{amsmath} 
\usepackage{bm} 
\usepackage{amsfonts} 
\usepackage{graphicx} 
\usepackage{float} 
\usepackage{caption} 
\usepackage{bigstrut} 
\usepackage{tabularx} 
\usepackage{cancel} 
\usepackage{subfig} 
\usepackage{xcolor}
\usepackage{soul}
\usepackage{upgreek}
\usepackage{slashed} 
\usepackage{epstopdf} 

\usepackage[subnum]{cases} 
  {\color{red}}%
  {}
\newcommand\nmax{n_{\text{max}}}
\newcommand\Shalf{\Sigma_{1/2}}
\newcommand\Shalfm{\Sigma_{-1/2}}
\newcommand\SSn{\Sigma_{S,n}}
\newcommand\Sc{\Sigma_X}

\title[MHD stability of magnetars II]{Magnetohydrodynamic stability of magnetars in the ultrastrong field regime II: The crust}

\author[P. B. Rau and I. Wasserman]{
Peter B. Rau$^{1}$\thanks{E-mail: prau@uw.edu}
and Ira Wasserman$^{2}$
\\
$^{1}$Institute for Nuclear Theory, University of Washington, Seattle, WA, U.S.A.
\\
$^{2}$Cornell Center for Astrophysics and Planetary Science, Cornell University, Ithaca, NY, U.S.A.
}

\date{}

\pubyear{2022}

\begin{document}
\label{firstpage}
\pagerange{\pageref{firstpage}--\pageref{lastpage}}
\maketitle


\begin{abstract}
We study the stability of Hall MHD with strong magnetic fields in which Landau quantization of electrons is important. We find that the strong-field Hall modes can be destabilized by the dependence of the differential magnetic susceptibility on magnetic field strength. This hydrodynamic instability, thermodynamic in origin and stabilized by magnetic domain formation, is studied using linear perturbation theory. It is found to have typical growth time of order $\lesssim 10^3$ yrs, with the growth time decreasing as a function of wavelength of the perturbation. The instability is self-limiting, turning off following a period of local field growth by a few percent of the initial value. Finite temperature is also shown to limit the instability, with sufficiently high temperatures eliminating it altogether. Alfv\'{e}n waves can show similar unstable behaviour on shorter timescales. We find that Ohmic heating due to the large fields developed via the instability and magnetic domain formation is not large enough to account for observed magnetar surface temperatures. However, Ohmic heating is enhanced by the oscillatory differential magnetic susceptibility of Landau-quantized electrons, which could be important to magneto-thermal simulations of neutron star crusts.
\end{abstract}


\section{Introduction}

Magnetars are a class of strongly-magnetized neutron stars with typical surface fields $\sim10^{14}$--$10^{15}$ G distinguished by their unique X-ray pulsations and repeating bursts of gamma rays~\citep{Mereghetti2015,Turolla2015,Kaspi2017}. One class of proposed mechanisms behind the bursts and more powerful but rarer giant flares involves magnetohydrodynamic (MHD) instabilities inside the magnetar leading to sudden release of energy into the surrounding magnetosphere through the excitation of Alfv\'{e}n waves along the magnetic field lines anchored to the stellar crust~\citep{Thompson1995,Heyl2005}. A second class involves a slower build-up of magnetospheric twisting leading to reconnection and the creation of a magnetically-confined pair plasma~\citep{Lyutikov2003,Lyutikov2006}. The exact mechanism for triggering these instabilities or generating the magnetospheric twists is not known, but both are associated with magnetic field evolution and crust motions or mechanical failures. As significant cracking and slipping of the crust is forbidden by the high pressures inside a neutron star~\citep{Jones2003,Levin2012}, the failure mechanism of the crust cannot be analogous to earthquakes. A more plausible possibility is that the crust undergoes plastic deformation once the magnetic stresses exceed a particular yielding value-- this has been investigated in the context of thermoplastic waves~\citep{Beloborodov2014} and Hall wave avalanches~\citep{Li2016} which have been suggested as the source of magnetar outbursts. MHD instabilities are also invoked in many classes of fast radio burst (FRB) models with magnetar progenitors (e.g.,~\citet{Metzger2019,Lu2020} and references therein), where they are responsible for excitation of Alfv\'{e}n waves which travel into the magnetosphere or  for releasing ejecta during flaring events.

An additional aspect of magnetar crust physics that is not fully understood is the anomalously high surface temperatures inferred from the thermal luminosity of many magnetars. They are observed to be significantly hotter than other isolated neutron stars, with typical luminosities $\mathscr{L}\approx10^{35}$ erg/s, roughly two orders of magnitude larger than those of standard rotation-powered pulsars. Many detailed simulations of their magneto-thermal evolution, including transport accounting for Landau quantization of fermions~\citep{Hernquist1984,Potekhin1999a,Potekhin2001} and different accreted envelopes (oceans) have been performed to try to understand this problem.~\citet{Pons2009} found increased heating in magnetars because of enhanced Joule heating from decay of crustal currents at strong magnetic field strengths.~\citet{Vigano2013} found that the high luminosities of magnetars could be explained by high conductivity light element envelopes. Simulations by~\citet{Potekhin2018} including transport by Landau-quantized electrons were able to explain the temperatures of about half the known magnetars, but discrepancies between the theoretical luminosity prediction and observation of around an order of magnitude remain for the other half. One possible explanation for this disagreement is an unaccounted-for heat source from dissipation of the magnetic field. A phenomenological calculation by~\citet{Kaminker2006} of interior heating in a magnetar argued that a heat source located within the outer crust of the star would be consistent with the observed high surface temperatures, as heat energy from a source deeper in the star would be cooled by neutrino emission before warming the surface.~\citet{Beloborodov2016} reviewed four different heating mechanisms that could be responsible for these hotter temperatures, but found none that were completely satisfactory in explaining both the observed surface luminosities \textit{and} magnetar lifetimes $\approx10^4$ yrs.~\citet{Akgun2018} proposed that Ohmic dissipation of currents in weakly-conducting magnetar envelopes could be sufficient to explain observed surface temperatures. More recently,~\citet{Chamel2021} have proposed electron capture and pycnonuclear fusion reactions triggered by magnetic field evolution within the crust as a viable heating source to explain the theory-observation discrepancy. 

In an idealized neutron star crust, only the electrons (and in the inner crust, the nonconducting superfluid neutrons) are free to move. The electrons thus carry the electric current, and the magnetic field moves with them with respect to the neutralizing nuclear lattice. Under these conditions, the ideal magnetohydrodynamics (MHD) applicable in the core of a neutron star must be supplanted with \textit{Hall MHD}, and the Hall term included in Ohm's Law. Hall MHD and Hall equilibria, the equilibrium field configurations when Hall MHD governs the evolution of the field, have been studied for neutron star crusts in many previous calculations (e.g.,~\citet{Jones1988,Goldreich1992,Rheinhardt2002,Cumming2004,
Gourgouliatos2013,Gourgouliatos2015,Lyutikov2015,Li2016}). A notable difference between ideal MHD and Hall MHD is that the latter lacks a (canonical) energy principle to determine stability: this was shown for electron MHD by~\citet{Lyutikov2013}, and this conclusion can be extended to the closely-related Hall MHD. Numerical MHD  simulations including the Hall term have demonstrated that, by generating small-scale spatial structures, it greatly enhances dissipation of magnetic field energy during the early stages of a neutron star's lifetime~\citep{Vigano2013}.

In~\citet{Rau2021} (henceforth Paper I), we studied MHD stability in ultrastrong magnetic fields relevant to magnetar cores using ideal MHD and a canonical energy principle~\citep{Friedman1978a,Glampedakis2007}. We included the magnetic field-dependence of the internal energy due to Landau quantization of fermions, showing how this can lead to a fast-growing, but spatially limited, instability. In this paper, we extend that study to magnetar crusts, replacing our ideal MHD formalism with Hall MHD. This is the first study including Landau quantization effects in Hall MHD. As mentioned, we \textit{cannot} use the canonical energy principle computed in Paper I, Eq.~8 to analyse stability in the crust. Instead we use perturbation theory to look for a strong-field Hall MHD instability. This is done through a linear mode analysis with coupling to the crust following~\citet{Cumming2004}. We first look at this instability at zero temperature, then include finite temperature and examine how temperature can limit the instability. In a subsequent paper we will examine Hall MHD including Landau quantization numerically to examine nonlinear effects. One of our goals is to examine the effect that Landau quantization has on stability and the potential to trigger magnetar outbursts.

In Section~\ref{sec:CrustThermo} we briefly outline the thermodynamics of a strongly magnetized outer crust, referring to Paper I where appropriate. The results in this section are used later in the stability analysis. Section~\ref{sec:HallMHD} describes Hall MHD and its linear perturbative modes, showing the mechanisms by which unstable Hall waves (whistler modes) could develop in the crust. In Section~\ref{sec:NonzeroTUnstablePS} we study the temperature dependence of this instability and determine the potentially unstable region in field--density--temperature parameter space. In Section~\ref{sec:PhysicalImplications} we examine the modification to Ohmic dissipation due to Landau quantization of electrons, and estimate the Ohmic heating resulting from the dissipation of the unstably growing magnetic field during magnetic domain formation. The astrophysical relevance of the instability is discussed in that section and in the conclusion. We work in Gaussian units and employ the Einstein summation convention using Latin letters as spatial indices $i=1,2,3$.

\section{Thermodynamics of a strongly magnetized crust}
\label{sec:CrustThermo}

We consider a neutron star crust of nuclei and electrons. In the inner crust, we make the simplifying assumption that the dripped neutron superfluid moves with the nuclear lattice-- this is equivalent to assuming strong entrainment between the superfluid neutrons and the nuclei~\citep{Chamel2017a}. To determine the MHD stability, we must compute certain thermodynamic derivatives which are given below. This section is organized similarly to, and with analogous content to, Section~4 of Paper I.

The total grand potential density $\Omega$ for the crust must include contributions for the Landau-quantized electron gas, the magnetic pressure, the nuclear lattice and the dripped neutron superfluid. As we will not be overly concerned with the latter two contributions, we can write $\Omega$ as
\begin{equation}
\Omega(\mu_{\text{e}},n_{\text{b}},Y_{\text{N}},w,A,Z,B,E^s_{ij})=-P_{\text{e}}(\mu_{\text{e}},B)+\frac{B^2}{8\pi}+\Omega_{\text{b}}(n_{\text{b}},Y_{\text{N}},w,A,Z,E^s_{ij}),
\label{eq:GrandPotential}
\end{equation}
where $\mu_{\text{e}}$ is the electron chemical potential, $n_{\text{b}}$ is the total number density of baryons including those in the lattice nuclei and the dripped neutrons, $Y_{\text{N}}$ is the fraction of baryons within the nuclei, $w$ is the volume fraction of each unit cell occupied by the nucleus, $A$ and $Z$ are the atomic and mass numbers of the nuclei, $E^s_{ij}$ is the (deviatoric) shear strain tensor and $B$ is the magnetic field magnitude. $P_{\text{e}}$ is the electron pressure given in e.g., Paper I, Eq.~(40), and $\Omega_{\text{b}}$ is the grand potential associated with the baryons i.e., the nuclei and the dripped neutrons, including lattice energy, surface energy terms, elastic deformation, etc., which typically appear in treatments of the crust (e.g.,~\citet{Baym1971a,Baym1971}). The dripped neutron contributions are excluded in the outer crust. Though the natural independent variables for $\Omega_{\text{b}}$ would be the chemical potentials associated with the lattice nuclei and dripped neutrons, we assume that we can perform a change of variable and write $\Omega_{\text{b}}$ in terms of the number densities as is often done in equation of state (EOS) calculations. The vacuum Euler--Heisenberg Lagrangian, included in $\Omega$ in Paper I, is not included in this paper as its effects on the stability are small; its inclusion would add an additional term to the vacuum field pressure term $B^2/(8\pi)$. We have neglected temperature in this expression, as its main role at temperatures relevant to most of the lifetime of a magnetar is to regulate divergences in thermodynamic functions at filled Landau levels. If finite temperature is included, Eq.~(\ref{eq:GrandPotential}) and the following equations can be easily generalized to include it, though it adds unnecessary complication for our purposes.

Charge neutrality requires that the electron number density $n_{\text{e}}$ equals $Yn_{\text{b}}$ where the proton fraction is $Y=Y_{\text{N}}Z/A$, but this must be imposed after taking required partial derivatives of $\Omega$ or the internal energy density $u$. The latter quantity is given by (noting that temperature is neglected and hence $u$ and the Helmholtz free energy density are equal)
\begin{equation}
u(n_{\text{e}},n_{\text{b}},Y_{\text{N}},w,A,Z,B,E^s_{ij})=u_{\text{e}}(n_{\text{e}},B)+\frac{B^2}{8\pi}+u_{\text{b}}(n_{\text{b}},Y_{\text{N}},w,A,Z,E^s_{ij}),
\end{equation}
where $u_{\text{e}}$ and $u_{\text{b}}$ are the electron and baryonic (nuclei plus dripped neutron) internal energy density respectively. $u$ and $\Omega$ are related by the Legendre transformation
\begin{equation}
u=\Omega+n_{\text{e}}\mu_{\text{e}}+n_{\text{b}}\mu_{\text{b}},
\end{equation}
where $\mu_{\text{b}}=\partial u/\partial n_{\text{b}}|_{Y_{\text{N}},w,A,Z,E_{ij}^s}$ is the chemical potential of the baryons.

The first law of thermodynamics for both $u$ and $\Omega$ are used to compute thermodynamic derivatives appearing in the MHD equations. Using arguments from Paper I Section 4, we have $H=4\pi\partial u/\partial B|_{n_{\text{e}}}=4\pi\partial \Omega/\partial B|_{\mu_{\text{e}}}$, and can define $M\equiv-\partial u_{\text{e}}/\partial B|_{n_e}=-u_B$ where $M$ is the usual definition of the magnetization and $u_B$ is the notation used in Paper I. We then have
\begin{equation}
\text{d}M=\left.\frac{\partial^2P_{\text{e}}}{\partial B^2}\right|_{\mu_{\text{e}}}\text{d}B+\frac{\partial^2P_{\text{e}}}{\partial B\partial\mu_{\text{e}}}\text{d}\mu_{\text{e}}=\left[\left.\frac{\partial^2P_{\text{e}}}{\partial B^2}\right|_{\mu_{\text{e}}}-\left(\left.\frac{\partial^2 P_{\text{e}}}{\partial\mu_{\text{e}}^2}\right|_B\right)^{-1}\left(\frac{\partial^2P_{\text{e}}}{\partial B\partial\mu_{\text{e}}}\right)^2\right]\text{d}B+\left(\left.\frac{\partial^2 P_{\text{e}}}{\partial\mu_{\text{e}}^2}\right|_B\right)^{-1}\frac{\partial^2P_{\text{e}}}{\partial B\partial\mu_{\text{e}}}\text{d}n_{\text{e}}\equiv\chi_n\text{d}B+\mathcal{M}_n\text{d}n_{\text{e}},
\label{eq:dM}
\end{equation}
$\chi_n=-u_{BB}$ and $\mathcal{M}_n=-u_{Bn}$ in the notation of Paper I, with the former being the differential magnetic susceptibility at fixed $n_{\text{e}}$. Eq.~(\ref{eq:dM}) is used to conveniently compute $\chi_n$ and $\mathcal{M}_n$ in terms of partial derivatives of $P_{\text{e}}$. We will also need partial derivatives of $P_{\text{e}}$, in particular
\begin{equation}
\left.\frac{\partial P_{\text{e}}}{\partial B}\right|_{n_{\text{e}}}=\left.\frac{\partial P_{\text{e}}}{\partial B}\right|_{\mu_{\text{e}}}-n_{\text{e}}\left(\left.\frac{\partial^2 P_{\text{e}}}{\partial\mu_{\text{e}}^2}\right|_B\right)^{-1}\frac{\partial^2P_{\text{e}}}{\partial B\partial\mu_{\text{e}}}=M-n_{\text{e}}\mathcal{M}_n.
\label{eq:dPdB}
\end{equation}
The expressions for the partial derivatives of $P_{\text{e}}$ with respect to $B$ and $\mu_{\text{e}}$ can be found in the online supplemental material for Paper I. The functional behaviour of $u_{BB}=-\chi_n$ is identical to $u_{BB}$ defined there but here it only includes the electrons, and $u_{Bn}=-\mathcal{M}_n$ is defined analogously to $u_{\rho B}$, though they are not identical. We treat the electrons as a free Fermi gas, ignoring the effects of the crystal lattice and electron band structure on their Landau quantization (see~\citet{Shoenberg1984} for how band structure modifies this picture).

\section{Perturbative Hall MHD modes in strong magnetic fields}
\label{sec:HallMHD}

The MHD stability of the crust must be studied in a different framework than in the core. This is because in the crust the electric current is due to the electrons moving relative to the (approximately) stationary, neutralizing nuclear lattice. For sufficiently weak fields, the ions are effectively decoupled from the magnetic field, which moves with the electrons~\citep{Cumming2004}. Ohm's Law thus must be modified to include the Hall term (e.g.,~\citet{Goldreich1992}):
\begin{equation}
E_i=-\frac{1}{c}\epsilon_{ijk}v_jB_k+\frac{1}{n_{\text{e}}ec}\epsilon_{ijk}J^{j}_eB^k+\frac{1}{\sigma}J_{e,i}=-\frac{1}{c}\epsilon_{ijk}v^{\text{e}}_jB_k+\frac{1}{\sigma}J_{e,i}.
\label{eq:OhmsLaw}
\end{equation}
$E_i$ is the electric field, $\epsilon_{ijk}$ is the Levi-Civita tensor, $B^i$ is the magnetic field, and $v_{\text{e}}^i$ and $v^i$ are the velocity of the electrons and nuclear lattice respectively. $\sigma$ is the conductivity, which is assumed isotropic for simplicity here, but in strong magnetic fields this is not true in general. MHD when this is used as Ohm's Law is known as Hall MHD. Additional terms accouting for e.g., the electron inertia and the electron pressure gradient~\citep{Biskamp2000,Cramer2001} are neglected. The (free) current density follows from Amp\`{e}re's Law
\begin{equation}
J^i_e=e(Zn_{\text{N}}v^i-n_{\text{e}}v_{\text{e}}^i)=\frac{c}{4\pi}\epsilon^{ijk}\nabla_jH_k,
\label{eq:AmperesLaw}
\end{equation}
where $Z$ the atomic number of the nuclei and $n_{\text{N}}$ the number density of nuclei. We assume local charge neutrality $Zn_{\text{N}}=n_{\text{e}}$.

First consider the case where the nuclear lattice is fixed $v^i=0$. The magnetic induction equation $\partial_tB^i=-c\epsilon^{ijk}\nabla_jE_k$ becomes
\begin{equation}
\partial_tB^i=-\frac{c}{4\pi e}\epsilon^{ijk}\nabla_j\left(\frac{1}{n_{\text{e}}}\epsilon_{k\ell m}B^m\epsilon^{\ell n p}\nabla_nH_p\right)-\frac{c^2}{4\pi}\epsilon^{ijk}\nabla_j\left(\frac{1}{\sigma}\epsilon_{k\ell m}\nabla^{\ell}H^m\right).
\end{equation}
Since the Ohmic dissipation timescale is much longer than the Hall timescale for magnetars~\citep{Goldreich1992}, we ignore Ohmic dissipation in our analytic calculation. The relevant magnetic induction equation is thus 
\begin{equation}
\partial_tB^i=-\frac{c}{4\pi e}\epsilon^{ijk}\nabla_j\left(\frac{1}{n_{\text{e}}}\epsilon_{k\ell m}B^m\epsilon^{\ell n p}\nabla_nH_p\right)=\epsilon^{ijk}\epsilon_{k\ell m}\nabla_{j}\left(v_{\text{e}}^{\ell}B^m\right).
\label{eq:MagneticInductionElectronMHD}
\end{equation}
This says that the magnetic field is frozen to the electron fluid. Magnetic fields that are static solutions to this equation i.e., $\partial_tB^i=0$, are Hall equilibria. 


As mentioned in the introduction, there is no canonical energy-based stability criterion for Hall MHD~\citep{Lyutikov2013}. Instead, we examine the local stability of the low-frequency oscillation modes that are perturbations of Eq.~(\ref{eq:MagneticInductionElectronMHD}): the Hall modes/waves, also referred to as whistler modes~\citep{Cramer2001}. These are a class of low-frequency, circularly-polarized MHD modes. We first consider the case where the nuclear lattice is decoupled from the electrons, which is the original case considered by~\citet{Goldreich1992}. We then consider the modifications to the Hall mode dispersion due to the coupling to the nuclear lattice via the Lorentz force acting on the nuclei i.e. the relaxation of the $v^i=0$ approximation-- this was first considered by~\citet{Cumming2004}.

\subsection{No coupling to lattice}

We take the Eulerian perturbation of Eq.~(\ref{eq:MagneticInductionElectronMHD}), assuming uniform background quantities $B^i$, $v_{\text{e}}$ (which is zero for uniform $B^i$), $n_{\text{e}}$, $M$, etc. This assumption is made for all calculations in this section. The perturbation of $H_k$ is
\begin{equation}
\delta H_k = (1-4\pi M/B)\delta B_k-4\pi(\chi_n-M/B)\hat{B}_k(\hat{B}^m\delta B_m)-4\pi \mathcal{M}_n\hat{B}_k\delta n_{\text{e}},
\label{eq:deltaH}
\end{equation}
where we employ $\mathcal{M}_n\equiv\partial^2u_{\text{e}}/\partial B\partial n_{\text{e}}$. We also assume incompressible perturbations and uniform $n_{\text{e}}$, so $\delta n_{\text{e}}=0$. Assuming harmonic spatial-temporal dependence $\delta B^k\propto\exp(ik_jx^j-i\omega t)$, we obtain
\begin{equation}
-i\omega\delta B^i=\frac{c(k_jH^j)}{4\pi en_{\text{e}}}\epsilon^{ijk}k_j\delta B_k-\frac{c(k_jB^j)}{en_{\text{e}}}\left(\chi_n-\frac{M}{B}\right)(\hat{B}_m\delta B^m)\epsilon^{ijk}k_j\hat{B}_k,
\end{equation}
where $\hat{B}^i=B^i/B$. Contract this with $\epsilon_{\ell ni}\hat{k}^n$, where $\hat{k}_n=k_n/k$, then insert the result back into the original equation to eliminate $\epsilon^{ijk}k_j\hat{B}_k$. Contracting the resulting expression with $\hat{B}_i$ gives
\begin{equation}
\omega=\pm\frac{ck|k_jB^j|}{4\pi en_{\text{e}}}\sqrt{\frac{H^2}{B^2}-4\pi\frac{H}{B}\left(\chi_n-\frac{M}{B}\right)\sin^2\theta_B}\approx\pm\omega_{\text{H}}\sqrt{1-4\pi\chi_n\sin^2\theta_B},
\label{eq:UncoupledHallMode}
\end{equation}
where $\omega_{\text{H}}$ is the normal ($B=H$) Hall frequency, $\cos\theta_B\equiv \hat{k}_j\hat{B}^j$, and we used that $H\approx B$ to within a few percent even for fields above the quantum critical field (e.g., Figure 4(a) of Paper I), implying $|M|/B\ll 1$. This mode is clearly unstable if the term under the square root is negative, and as shown in Paper I, this can clearly be true at locations in $B$--$n_{\text{e}}$ parameter space where additional Landau levels are filled. If $\hat{B}_m\delta B^m=0$, Eq.~(\ref{eq:UncoupledHallMode}) reduces to $\omega=\omega_{\text{H}}\frac{H}{B}$: in this case, the modes are stable for a spatially uniform background.

In the unstable regions of parameter space, the unstable Hall modes have an approximate growth time (using Eq.~(\ref{eq:UncoupledHallMode}) with $B\approx H$)
\begin{equation}
\tau=\frac{1}{\text{Im}(\omega)}\sim 6\times10^3|\sec\theta_B|\left(\frac{10^{15}\text{ G}}{B}\right)\left(\frac{n_{\text{e}}}{10^{-4}\text{ fm}^{-3}}\right)\left(\frac{R_c^{-1}}{k}\right)^2\left(\frac{1}{|1-4\pi\chi_n\sin^2\theta_B|}\right)^{1/2}\text{ yrs},
\label{eq:UncoupledGrowthTime}
\end{equation}
where $R_c=10^{5}$ cm is the approximate thickness of the crust. This instability should be contrasted with ``density-shear instability''~\citep{Wood2014} of Hall MHD which requires a gradient in $n_{\text{e}}$ and shear in $v_{\text{e}}$, or of the resistive tearing instability  discussed in the context of Hall MHD, often under the name ``Hall drift instability'', by~\citet{Rheinhardt2002} and later by~\citet{Rheinhardt2004},~\citet{Cumming2004} and~\citet{Pons2010}. The instability in this paper will occur in any sufficiently strongly magnetized crust at up to hundreds of spatially-restricted regions where $1-4\pi\chi_n\sin^2\theta_B<0$. However, it is much slower than the MPR-type instability in Paper I due to the Hall phase velocity $\omega_{\text{H}}/k$ being much slower than the Alfv\'{e}n velocity, which set the characteristic instability timescales in this paper and Paper I respectively. We note that regions where $\chi_n>1/(4\pi)$ are \textit{thermodynamically} unstable to magnetic domain formation and thus not in an equilibrium configuration, though the timescale to equilibrate could exceed the timescale of the hydrodynamic instability discussed above and in the remainder of this section; this is discussed in further detail in Section~\ref{sec:DomainFormation}.

\subsection{Coupling to lattice}
\label{sec:ModeWithLatticeCoupling}

Motion of the nuclear lattice is described by the Euler equation with strong-field magnetization terms (Eq. (3) of Paper I). In the solid crust we additionally include shear stresses, since for magnetar-strength magnetic fields in the crust, the elasticity must be included to balance the Lorentz force exerted by the magnetic field. Using Eq.~(\ref{eq:GrandPotential}) and $P\equiv P_{\text{e}}+\Omega_{\text{b}}$, we can write the Euler equation as
\begin{equation}
\rho(\partial_t v_i+v^j\nabla_jv_i)+\nabla_iP+\rho\nabla_i\Phi=\nabla^j\left[\frac{B^2}{8\pi}g_{ij}-\frac{1}{4\pi}H_kB^kg_{ij}+\frac{1}{4\pi}H_iB_j\right]+\nabla^j\sigma_{ij}\equiv\nabla^jT^B_{ij}+\nabla^j\sigma_{ij}.
\label{eq:LatticeEulerEq}
\end{equation}
where $g_{ij}$ is the metric tensor and the (deviatoric) shear stress tensor is
\begin{equation}
\sigma_{ij}=\check{\mu}\left(\nabla_i X_j+\nabla_j X_i-\frac{2}{3}\nabla_k X^kg_{ij}\right)=2\check{\mu}E_{ij}^{s}.
\label{eq:ShearStressElastic}
\end{equation}
$X^i$ is the local displacement of the crust from its relaxed state, $\check{\mu}$ is the shear modulus, and $E_{ij}^{s}$ is the (deviatoric) shear strain tensor. The shear modulus of a Coulombic nuclear lattice is~\citep{Strohmayer1991}
\begin{equation}
\check{\mu}\approx 0.1925e^2Z^{2/3}n_{\text{e}}^{4/3}=2.4\times10^{28}\left(\frac{Z}{40}\right)^{2/3}\left(\frac{n_{\text{e}}}{10^{35}\text{ cm}^{-3}}\right)^{4/3}\text{ g s}^{-2}\text{ cm}^{-1},
\end{equation}
where we ignore thermal corrections. For comparison, the electron pressure is $P_{\text{e}}\approx (3\pi^2)^{1/3}\hbar c n_{\text{e}}^{4/3}/4\approx 53(40/Z)^{2/3}{\check\mu}$ if many Landau levels are occupied.

We have three quantities describing the perturbations: the Eulerian perturbation of the magnetic field $\delta B^i$, and the Lagrangian displacement fields for the electrons and lattice $\xi^i_{\text{e}}$ and $\xi^i$ respectively. $\delta v^i=\partial_t\xi^i$ and $\delta v^i_{\text{e}}=\partial_t\xi^i_{\text{e}}$ in the limit of zero background velocities. The perturbed forms of three equations,  Eq.~(\ref{eq:AmperesLaw},\ref{eq:MagneticInductionElectronMHD},\ref{eq:LatticeEulerEq}), determine these perturbations, so two of the perturbed quantities can be eliminated.

Taking the Eulerian perturbation of Eq.~(\ref{eq:AmperesLaw}) and using that $v^i=0=v^i_{\text{e}}$ in the background,
\begin{equation}
en_{\text{e}}(\delta v^i-\delta v^i_{\text{e}})=\frac{c}{4\pi}\epsilon^{ijk}\nabla_j\delta H_k.
\label{eq:PerturbedAmperesLaw}
\end{equation}
Assuming $\xi^i,\xi^i_{\text{e}}\propto\exp(ik_jx^j-i\omega t)$ and using Eq.~(\ref{eq:deltaH}) and $\delta n_{\text{e}}=-n_{\text{e}}\nabla_j\xi^j_{\text{e}}$ for uniform $n_{\text{e}}$, we obtain
\begin{equation}
-i\omega en_{\text{e}}(\xi^i-\xi^i_{\text{e}})=\frac{ic}{4\pi}\left[\frac{H}{B}\epsilon^{ijk}k_j\delta B_k-4\pi\left(\chi_n-\frac{M}{B}\right)\epsilon^{ijk}k_j\hat{B}_k(\hat{B}_m\delta B^m)+4\pi n_{\text{e}}\mathcal{M}_n(k_m\xi^m_{\text{e}})\epsilon^{ijk}k_j\hat{B}_k\right].
\end{equation}
Contracting this equation with $k_i$ implies $k_j\xi^j=k_j\xi^j_{\text{e}}$; using this condition to eliminate $k_j\xi^j_{\text{e}}$ gives an expression for $\xi^i_{\text{e}}$ in terms of $\xi^i$ and $\delta B^i$ alone:
\begin{equation}
\xi^i_{\text{e}}=\xi^i+\frac{c\mathcal{M}_n}{e\omega}(k_m\xi^m)\epsilon^{ijk}k_j\hat{B}_k+\frac{c}{4\pi en_{\text{e}}\omega}\left[\frac{H}{B}\epsilon^{ijk}k_j\delta B_k-4\pi\left(\chi_n-\frac{M}{B}\right)\epsilon^{ijk}k_j\hat{B}_k(\hat{B}_m\delta B^m)\right].
\label{eq:XiE}
\end{equation}
For uniform background field the perturbed induction equation, using as $E^i$ the second form of Eq.~(\ref{eq:OhmsLaw}), we have
\begin{equation}
\partial_t\delta B^i=B^k\nabla_k\delta v_{\text{e}}^i-B^i\nabla_k\delta v_{\text{e}}^k=B^k\nabla_k\partial_t\xi^i_{\text{e}}-B^i\nabla_k\partial_t\xi^k_{\text{e}}.
\end{equation}
Eliminating $\xi_{\text{e}}$ using Eq.~(\ref{eq:XiE}) gives
\begin{equation}
-i\omega\delta B^i=(k_jB^j)\omega\xi^i-B^i(k_j\xi^j)\omega+\frac{c(k_jB^j)}{e}\mathcal{M}_n(k_m\xi^m)\epsilon^{ijk}k_j\hat{B}_k+\frac{c(k_jH^j)}{4\pi en_{\text{e}}}\epsilon^{ijk}k_j\delta B_k+\frac{c(k_jB^j)}{en_{\text{e}}}\left(\chi_n-\frac{M}{B}\right)(\hat{B}_m\delta B^m)\epsilon^{ijk}k_j\hat{B}_k
\label{eq:PerturbedInductionEq}
\end{equation}

Taking the Eulerian perturbation of Eq.~(\ref{eq:LatticeEulerEq}) and using the magnetohydrostatic form of Eq.~(\ref{eq:LatticeEulerEq}), ignoring gravity and gradients of background quantities, we find
\begin{equation}
\partial_t^2\xi_i+\frac{1}{\rho}\nabla_i\delta P=\frac{1}{\rho}\nabla^j\delta T^B_{ij}+\frac{1}{\rho}\nabla^j\delta \sigma_{ij},
\label{eq:PrelimPerturbedLatticeEulerEq}
\end{equation}
where
\begin{align}
\delta T^B_{ij}=\frac{1}{4\pi}\left(B_k\delta B^k-B_k\delta H^k-H_k\delta B^k\right)g_{ij}+\frac{1}{4\pi}\left(B_j\delta H_i+H_i\delta B_j\right),
&&
\delta\sigma_{ij}=\check{\mu}\left(\nabla_i \xi_j+\nabla_j \xi_i-\frac{2}{3}\nabla_k \xi^k\delta_{ij}\right),
\end{align}
assuming spatially uniform $\check{\mu}$. Eq.~(\ref{eq:PrelimPerturbedLatticeEulerEq}) can then be written as
\begin{align}
(k^2\check{\mu}-\rho\omega^2)\xi_i+\left(i\delta P+\frac{\check{\mu}}{3}(k_j\xi^j)\right)k_i={}&i\Bigg[\frac{(k_jH^j)}{4\pi}\delta B_i
-(k_jB^j)\left\{\left(\chi_n-\frac{M}{B}\right)(\hat{B}_m\delta B^m)+in_{\text{e}}\mathcal{M}_n(k_j\xi^j)\right\}\hat{B}_i
\nonumber
\\
{}&\quad-\left\{in_{\text{e}}\mathcal{M}_nB(k_j\xi^j)+\frac{H}{4\pi}(\hat{B}_j\delta B^j)-B\chi_n(\hat{B}_j\delta B^j) \right\}k_i\Bigg].
\label{eq:PerturbedLatticeEulerEq}
\end{align}
Using Eq.~(\ref{eq:PerturbedInductionEq},\ref{eq:PerturbedLatticeEulerEq}), we can find the dispersion relation for Hall MHD modes when the electron fluid and the nuclear lattice are coupled. 

We work in the incompressible limit $k_j\xi^j\to 0$. Within this approximation,  we find that the ratio of the component of $\xi^j$ along $k^j$ is $\sim v_A^2/c_s^2$ times the magnitude of $\xi^j$ perpendicular to $k^j$, where $v_A^2=B^2/4\pi\rho$ and $c_s^2$ are the squares of the Alfven and sound speeds, respectively. Roughly speaking, the incompressible approximation breaks down for all modes once the magnetic pressure exceeds the matter pressure, hence is only valid at sufficiently high densities within the crust. Define the Alfv\'{e}n frequency $\omega_{\text{A}}$ and shear mode frequency $\omega_s$ through
\begin{align}
\omega^2_{\text{A}}\equiv \frac{(k_jB^j)(k_jH^j)}{4\pi\rho}, &&
\omega^2_s\equiv \frac{\check{\mu}k^2}{\rho}.
\label{eq:AlfvenAndShearFrequency}
\end{align}
Eq.~(\ref{eq:PerturbedInductionEq}) and~(\ref{eq:PerturbedLatticeEulerEq}) now become
\begin{align}
-i\omega\delta B^i&{}=\frac{c(k_{\ell}H^{\ell})}{4\pi en_{\text{e}}}\epsilon^{ijk}k_j\delta B_k-\frac{c(k_jB^j)}{en_{\text{e}}}\left(\chi_n-\frac{M}{B}\right)\epsilon^{ijk}k_j\hat{B}_k(\hat{B}_m\delta B^m)+\omega(k^{\ell}B_{\ell})\xi^i,
\label{eq:PerturbedInductionEqIncompressible}
\\
(\omega_s^2-\omega^2)\xi_i{}&=i\frac{(k_jH^j)}{4\pi}\delta B_i
-i(k_jB^j)\left(\chi_n-\frac{M}{B}\right)(\hat{B}_m\delta B^m)\left(\hat{B}_i-\hat{k}_i(\hat{k}_j\hat{B}^j)\right).
\label{eq:PerturbedLatticeEulerEqIncompressible}
\end{align}
since $\xi^i$ cannot have a component along $k^i$, and where $\hat{B}^i-\hat{k}^i(\hat{k}_j\hat{B}^j)$ was used to replace $\hat{B}^i$ as it is the projection of $\hat{B}^i$ perpendicular to $k^i$.

Using Eq.~(\ref{eq:PerturbedLatticeEulerEqIncompressible}) to eliminate $\xi^i$ from Eq.~(\ref{eq:PerturbedInductionEqIncompressible})
, then contracting the result with $B^i$ and $\epsilon^{ijk}k_j\hat{B}_k$ gives
\begin{align}
-i\omega\left(\frac{\omega^2-\omega_s^2-\omega^2_{\text{A}}}{\omega^2-\omega_s^2}\right)\epsilon_{ijk}\hat{B}^i\hat{k}^j\delta B^k{}&=-\frac{ck(k_{\ell}H^{\ell})}{4\pi en_{\text{e}}}(\hat{B}^j\delta B_j)+\frac{ck(k_{\ell}B^{\ell})}{en_{\text{e}}}\left(\chi_n-\frac{M}{B}\right)(\hat{B}^j\delta B_j)\left(1-(\hat{k}^j\hat{B}_j)^2\right),
\\
-i\omega\left[\frac{\omega^2-\omega_s^2-\omega^2_{\text{A}}(1-\alpha)}{\omega^2-\omega_s^2}\right](\hat{B}^j\delta B_j){}&=\frac{ck(k_{\ell}H^{\ell})}{4\pi en_{\text{e}}}\epsilon_{ijk}\hat{B}^i\hat{k}^j\delta B^k.
\end{align}
where we've defined $\hat{k}^j\hat{B}_j=\cos\theta_B$ and
\begin{equation}
\alpha\equiv\frac{4\pi B}{H}\left(\chi_n-\frac{M}{B}\right)\sin^2\theta_B\approx 4\pi\chi_n\sin^2\theta_B,
\label{eq:alphaFactor}
\end{equation}
where the approximate form is valid because $H\approx B$ and $|M|/B\ll 1$. Combining these equations we obtain the dispersion relation
\begin{equation}
\omega^2=\frac{\omega_{\text{H}}^2(\omega^2-\omega_s^2)^2(1-\alpha)}{\left(\omega^2-\omega_s^2-\omega^2_{\text{A}}\right)\left(\omega^2-\omega_s^2-\omega_{\text{A}}^2(1-\alpha)\right)}.
\label{eq:CoupledHallModeFull}
\end{equation}
In the $\omega_s\rightarrow\infty$ limit this recovers Eq.~(\ref{eq:UncoupledHallMode}) as expected, but in the $H\rightarrow B$ limit it does not recover Eq. (47) of~\cite{Cumming2004}. In the low frequency limit $|\omega^2|\ll\omega_s^2,\omega_{\text{A}}^2$, it becomes
\begin{equation}
\omega^2\approx\frac{\omega_{\text{H}}^2(1-\alpha)}{\left(1+\omega^2_{\text{A}}/\omega_s^2\right)\left(1+\omega_{\text{A}}^2(1-\alpha)/\omega_s^2\right)}.
\label{eq:CoupledHallModeLowFreq}
\end{equation}
which reproduces~\cite{Cumming2004} Eq. (48) (their Eq. (47) contains a typographical error).

Eq.~(\ref{eq:CoupledHallModeLowFreq}) supports an unstable Hall mode if $\alpha\approx 4\pi\chi_n\sin^2\theta_B>1$, which is the same source of instability as the uncoupled case in Eq.~(\ref{eq:UncoupledHallMode}). For typical magnetar field strengths and crust densities, $\omega_{\text{A}}$ can be much greater than $\omega_s$, so for a crust with finite rigidity the mode frequency is reduced and  the instability growth time is increased compared to the estimate made in Eq.~(\ref{eq:UncoupledGrowthTime}) for an infinitely rigid crust.

\begin{figure}
\center
\includegraphics[width=0.95\linewidth]{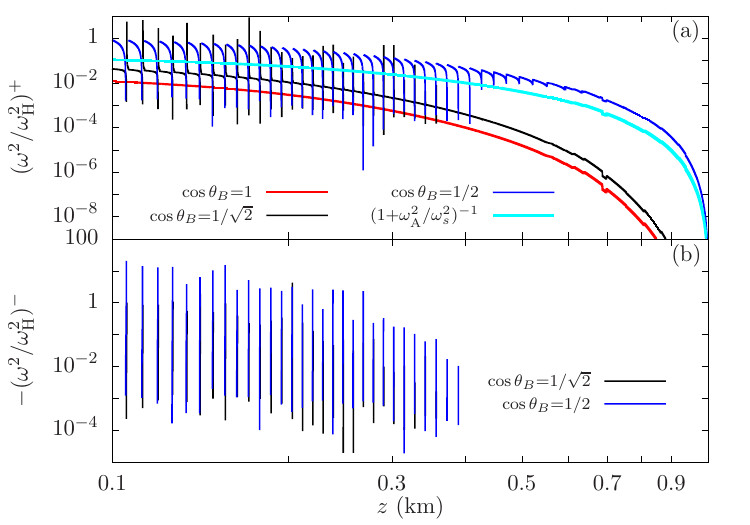}
\caption[Strong-field Hall MHD dispersion]{\label{fig:HallMHDDisp} Approximate (low frequency) dispersion relation for strong-field Hall MHD, normalized by $\omega_{\text{H}}^2$, as a function of position outward from the crust-core boundary $z$ and computing using Eq.~(\ref{eq:CoupledHallModeLowFreq}). Uniform field $B=10^{16}$ G and varying angles between the wave vector $k$ and magnetic field $B$ are shown. Note that the ``kinks'' not associated with Landau quantization originate from density discontinuities at transitions from nuclear layers. (a): Positive part. The standard crust-coupled Hall MHD result $\omega_{\text{H}}^2/(1+\omega_{\text{A}}^2/\omega_s^2)$~\citep{Cumming2004} is overlaid for comparison. (b): Negative of the negative part. Unstable regions are those with nonzero negative part.}
\end{figure}

The low-frequency approximation Eq.~(\ref{eq:CoupledHallModeLowFreq}) for the strong-field, crust-coupled Hall MHD modes is plotted in Figure~\ref{fig:HallMHDDisp} for uniform field $B=10^{16}$ G and a realistic crust profile. For the crust EOS we employ the BSk24 EOS~\citep{Pearson2018}. The crust used is that for a $1.4M_{\odot}$ star with radius $R=12.59$ km, crust thickness $R_c=1.04$ km and neutron drip line $0.48$ km from the surface. The significant magnetic forces that would act on the crust have been ignored in obtaining the background model, so these results  illustrate typical numerical values associated with these Hall modes rather than providing exact values. Panel (a) shows $\omega^2>0$ i.e., where the Hall modes are oscillatory, whereas panel (b) shows $\omega^2<0$ i.e., where they are unstable. The instability is possible in regions in $B$--$n_{\text{e}}$ parameter space where new Landau levels are populated. However, there is no instability if $\cos\theta_B$ is not sufficiently small, explaining the absence of the $\cos\theta_B=1$ curve from panel (b).

\begin{figure}
\center
\includegraphics[width=0.8\linewidth]{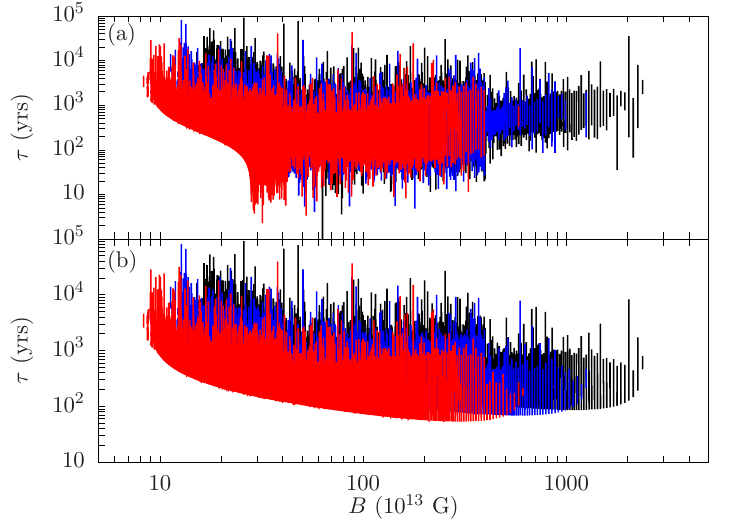}
\caption[Strong-field Hall MHD instability growth time]{\label{fig:HallMHDGrowthTime} Growth time $\tau$ of the instability associated with the strong-field Hall modes as a function of $B$. $\tau$ is computed using the low frequency approximate dispersion relation with coupling to the lattice, Eq.~(\ref{eq:CoupledHallModeLowFreq}) (panel (a)) and without coupling to the lattice, Eq.~(\ref{eq:UncoupledHallMode}) (panel (b)). The growth times for three different locations within the crust for a single representative wave number $k=(R_c/10)^{-1}=10^{-4}$ cm$^{-1}$ is shown. The curves are evaluated at $\mu_{\text{e}}=71$, $53$, $38$ MeV, correspondingly approximately to the electron density at $z=0.1$ (black), $0.25$ (blue) and $0.4$ km (red) respectively inside the crust. $\chi_n$ is evaluated at $T=4\times10^{7}$ K.}
\end{figure}

The growth time of the unstable Hall mode at a particular location within the star will change as the field evolves. Figure~\ref{fig:HallMHDGrowthTime} shows this evolution for different locations within the model crust, plotting Eq.~(\ref{eq:UncoupledGrowthTime}) with $\omega$ from Eq.~(\ref{eq:CoupledHallModeLowFreq}) used instead of Eq.~(\ref{eq:UncoupledHallMode}) in panel (a) and Eq.~(\ref{eq:UncoupledGrowthTime}) directly in panel (b). Note that the particular locations within the crust are distinguished by their electron chemical potentials and not $n_{\text{e}}$, which is a function of $B$ and changes up to a few percent as the field is varied. Typical growth times at the specified wave number $k=R_c/10=10^{-4}$ cm$^{-1}$ are $\sim 10^2$--$10^4$ yrs, which are comparable to magnetar lifetimes.

Importantly, the fluid becomes stable as the strongly-quantized threshold (all electrons in a single Landau level) is approached and $n\lesssim10$ Landau levels are occupied. This only occurs within the plotted range of fields for $\mu_{\text{e}}<24.33$ MeV, corresponding to regions of the crust with $z\gtrsim 0.45$ km. Increasing temperature (not shown) has a stabilizing effect as in Paper I, as it smooths the sharp transitions where the instability occurs and either increases $\tau$ or stabilizes the fluid. In contrast, decreasing the temperature allows the instability to persist at lower field strengths, lowering the apparent low-field cutoffs in the $z=0.1$ km and $z=0.5$ km growth times. The growth time is  very sensitive to $B$, and can change by orders of magnitude and even stabilize with a single percent change in $B$. However stable Hall mode propagation can change the local field value by small amounts and destabilize a region of the crust that was previously stable.

In the high frequency limit (the incompressible ideal MHD limit), noting that $\omega_{\text{A}}^2,|\omega^2|\gg\omega_s^2,\omega^2_{\text{H}}$, we find solutions
\begin{equation}
\omega^2=\omega^2_{\text{A}},~\omega^2_{\text{A}}(1-\alpha),
\end{equation}
which have $\hat{B}^i\delta B_i=0$ and $\hat{B}^i\delta B_i\neq0$ respectively. The first solution is (stable) Alfv\'{e}n modes, whereas the second solution can be both stable and unstable depending on $B$ and $\mu_{\text{e}}$ (or $n_{\text{e}}$). The unstable Alfv\'{e}n modes are the analog to the unstable modes discussed in Paper I; the assumption of incompressibility means that the term proportional to the cross derivative of $u$ with respect to $B$ and $n_{\text{e}}$ (in Paper I, the mass density $\rho$) is absent here. These modes thus have much faster growth times than the unstable Hall modes. Since we have ignored the displacement current term in Amp\`{e}re's Law, the resulting Alfv\'{e}n modes can be superluminal for sufficiently large $B^2/\rho$ as is possible in magnetar crusts; including the displacement field term, we instead obtain
\begin{equation}
\omega^2=\frac{\omega^2_{\text{A}}}{1+B^2/(4\pi\rho c^2)},~\frac{\omega_{\text{A}}^2(1-\alpha)}{1+B^2\cos^2\theta_B/(4\pi\rho c^2)}.
\end{equation}
At sufficiently low density, $B^2/4\pi$ exceeds $\rho$ times the sound speed squared  $c_s^2$ and the incompressible approximation fails. Since electrons dominate the pressure at low enough density, magnetic fields dominate once $n_{\text{e}}\approx 1.4\times 10^{34}B_{15}^{3/2}{\rm cm^{-3}}$. Note that approximately $\sqrt{3\pi/4e^2}=13.7$ Landau levels are filled at this density.

\subsection{Effect of finite conductivity}

It is interesting to ask whether Ohmic decay can damp the unstable modes discussed in the previous section. It can be incorporated into our analysis by retaining the  conductivity term in Eq.~(\ref{eq:OhmsLaw}). Following analogous manipulations to those in Section~\ref{sec:ModeWithLatticeCoupling}, we obtain the dispersion relation including finite conductivity for incompressible modes:
\begin{equation}
\omega^2=\frac{\omega_{\text{H}}^2(\omega^2-\omega_s^2)^2(1-\alpha)}{\left[\omega^2-\omega_s^2-\omega^2_{\text{A}}(1-\alpha)+i\gamma_{\text{O}}\omega^{-1}(\omega^2-\omega_s^2)\left(1-4\pi\frac{M}{B}-\frac{H}{B}\alpha\right)\right]\left[\omega^2-\omega_s^2-\omega^2_{\text{A}}+i\gamma_{\text{O}}\omega^{-1}(\omega^2-\omega_s^2)\left(1-4\pi\frac{M}{B}\right)\right]},
\label{eq:DispersionOhmic}
\end{equation}
where $\gamma_{\text{O}}\equiv k^2c^2/(4\pi\sigma)$ and $\alpha$ is as defined in Eq.~(\ref{eq:alphaFactor}). Ohmic decay alters the Hall mode in both the stable and unstable regimes. For very low frequencies we get
\begin{equation}
\left[\omega+\frac{i\gamma_{\text{O}}(1-\alpha)}{1+\omega_A^2(1-\alpha)/\omega_s^2}\right]\left(\omega+\frac{i\gamma_{\text{O}}}{1+\omega_{\text{A}}^2/\omega_s^2}\right)
=\frac{\omega_H^2(1-\alpha)}{[1+\omega_{\text{A}}^2(1-\alpha)/\omega_s^2](1+\omega_{\text{A}}^2/\omega_s^2)},
\label{eq:LowFreqOhmModes}
\end{equation}
instead of Eq. (32). In the infinitely rigid limit, $\omega_s^2\gg\omega_{\text{A}}^2$, Eq. ~(\ref{eq:LowFreqOhmModes}) has the general solution
\begin{equation}
\omega_{\pm}=\frac{1}{2}\left[-i\gamma_{\text{O}}(2-\alpha)\pm\sqrt{-\gamma_{\text{O}}^2(2-\alpha)^2+4(\gamma_{\text{O}}^2+\omega_H^2)(1-\alpha)}\right]
=\frac{1}{2}\left[-i\gamma_{\text{O}}(2-\alpha)\pm\sqrt{-\gamma_{\text{O}}^2\alpha^2+4\omega_H^2(1-\alpha)}\right],
\end{equation}
which are damped oscillations as long as $\left\vert\alpha+2\omega_H^2/\gamma_{\text{O}}^2\right\vert<(2\omega_H^2/\gamma_{\text{O}}^2)\sqrt{1+\gamma_{\text{O}}^2/\omega_H^2}$ which, for $\alpha>0$, is equivalent to $\alpha<{2}/[{1+\sqrt{1+\gamma_{\text{O}}^2/\omega_H^2}}\,]<1$. Clearly, modes are purely imaginary for $\alpha>1$; substitute $\omega_{\pm}=is_{\pm}$ to get
\begin{equation}
s_{\pm}=\frac{1}{2}\left[-\gamma_{\text{O}}(2-\alpha)\pm\sqrt{\gamma_{\text{O}}^2(2-\alpha)^2+4(\omega_H^2+\gamma_{\text{O}})^2(\alpha-1)}\right].
\label{eq:LowFreqOhmGrowth}
\end{equation}
Surprisingly, Ohmic dissipation does not thwart the instability: $s_+>0$ for any value of $\gamma_{\text{O}}$. For $\omega_H=0$ Eq.~(\ref{eq:LowFreqOhmGrowth}) gives
$s_-=-\gamma_{\text{O}}$, which is pure Ohmic decay, but $s_+=\gamma_{\text{O}}(\alpha-1)$. We have verified that this growing mode is found in the perfectly rigid
limit when solving Faraday's law with $E^i=J^i_e/\sigma$ along with $\epsilon^{ijk}\nabla_jH_k=4\pi J^i_e/c$ as long as $\alpha>1$. The Alfv\'{e}n modes are also damped: in the high frequency limit and taking $\omega=\omega_0-i\Gamma$ we find using Eq.~(\ref{eq:DispersionOhmic})
\begin{align}
\omega_0^2\approx\omega_{\text{A}}^2,~\omega_{\text{A}}^2(1-\alpha),
&&
\Gamma\approx\frac{\gamma_{\text{O}}}{2},~\frac{\gamma_{\text{O}}(1-\alpha)}{2}.
\end{align}

\subsection{Effect of plastic flow}
\label{sec:PlasticFlow}

The magnetic fields attainable in magnetar crusts are sufficiently strong to exceed the yield strength of the crust, resulting in its plastic deformation. The crust can be modeled (e.g.,~\citet{Beloborodov2014}) as an elasto-viscoplastic solid~\citep{Irgens2008}. Below the yield criterion, the crust behaves as an elastic material with shear stress tensor given by Eq.~(\ref{eq:ShearStressElastic}). The yield stress $\sigma_y$ of a neutron star crust has been estimated through numerical simulations to be
\begin{equation}
\sigma_y\approx 0.03143 e^2Z^{2/3}n_{\text{e}}^{4/3}\approx \check{\mu}E_y.
\end{equation}
Here $E_y\sim 0.1$ is the yield strain~\citep{Horowitz2009,Chugunov2010,Caplan2018} and we have ignored thermal corrections. The exact value of $E_y$ is not well known, and may be lower than the value used here (e.g.,~\citet{Baiko2018}). The transition to the plastic regime begins when $\sigma_{\text{M}}>\sigma_y$ where $\sigma_{\text{M}}$ is the von Mises stress
\begin{equation}
\sigma_{\text{M}}=\sqrt{\frac{3}{2}\sigma_{ij}\sigma^{ij}}.
\end{equation}
Beyond this point, the shear stress is determined using the Bingham--Maxwell model
\begin{equation}
\frac{1}{2\eta}\left[1-\frac{\sigma_y}{\sigma_{\text{M}}}\right]\theta(\sigma_{\text{M}}-\sigma_y)\sigma_{ij}+\frac{1}{2\check{\mu}}\partial_t\sigma_{ij}=\partial_tE^{s}_{ij},
\label{eq:ShearStressViscoplastic}
\end{equation}
where $\theta(x)$ is the step function and $\eta$ is the effective dynamic viscosity in the plastic flow regime. $\eta$ is poorly understood theoretically for neutron star crusts: estimates for its value based on studies of magnetar magneto-plastic field evolution are of order $10^{36}$--$10^{38}$ g cm$^{-1}$ s$^{-1}$~\citep{Lander2016,Lander2019}. A theoretical calculation of its (temperature-dependent) value found $\eta\sim10^{31}$ g cm$^{-1}$ s$^{-1}$~\citep{Kwang-Hua2018}, comparable to the effective value of $\approx7\times10^{31}$ g cm$^{-1}$ s$^{-1}$ employed by~\citet{Li2016}, though the previous references argue that this value is too low to support the persistent twisting of magnetic field lines anchored to the crust which is a leading mechanism for generating magnetar outbursts.

Assuming harmonic time dependence $\exp(-i\omega t)$ for the displacement, Eq.~(\ref{eq:ShearStressViscoplastic}) implies a linear relation between $\sigma_{ij}$ and $E^s_{ij}$ of
\begin{equation}
\sigma_{ij}=2\left[\frac{\check{\mu}}{1+i\check{\mu}/(\eta\omega)\left[1-\sigma_y/\sigma_{\text{M}}\right]\theta(\sigma_{\text{M}}-\sigma_y)}\right]\equiv 2\tilde{\mu}(\omega)E^s_{ij},
\end{equation}
so to determine the effect of crust plasticity on the Hall mode dispersion relation amounts to replacing $\check{\mu}$ in $\omega_s^2$ with the dynamic modulus $\tilde{\mu}(\omega)$.

To estimate the von Mises stress we start with the MHD force balance equation, which can be written as the divergence of a tensor $T_{ij}=T^B_{ij}-\sigma_{ij}-Pg_{ij}+T^g_{ij}$ where $T^g_{ij}$ is the gravitational field tensor; integrating implies that $T_{ij}$ is equal to a divergence free tensor, which is of course force-free. For the magnetic stress, we isolate the part  with forces due to local currents in the crust, which excludes any force-free contribution. Assuming that the gravitational stresses are not important, the trace free part of this magnetic stress tensor should roughly balance $\sigma_{ij}$. We hence estimate
\begin{equation}
\sigma_{\text{M}}\approx \frac{B^2}{4\pi},
\end{equation}
so the crust will be in the plastic flow regime for densities below the threshold
\begin{equation}
\rho_{\text{crit}}\approx\frac{m_{\text{n}}n_{\text{e}}}{Y}\approx 5\times 10^{12}\left(\frac{40}{Z}\right)^{1/2}\left(\frac{0.5}{Y}\right)\left(\frac{B}{10^{15}\text{ G}}\right)^{3/2}\left(\frac{0.1}{E_y}\right)^{3/4}\text{ g cm}^{-3},
\end{equation}
for neutron mass $m_{\text{n}}$. We note that $\rho_{\text{crit}}$ is greater than the density at which $B^2/(8\pi)\approx P_{\text{e}}$. For $\sigma_{\text{M}}\geq\sigma_y$, there are two limiting cases. Define $\overline{\eta}=\eta(1-\sigma_y/\sigma_{\text{M}})$ and an effective relaxation time $\tau=\overline{\eta}/\check{\mu}$~\citep{Landau1970}. In the solid limit $|\tau\omega|\gg 1$ we have $\tilde{\mu}(\omega)\approx\check{\mu}(1-i\check{\mu}/(\eta\omega))$, while in the fluid limit $|\tau\omega|\ll 1$ and $\tilde{\mu}(\omega)\approx-i\overline{\eta}\omega$.

For the $|\tau\omega|\gg 1$ case, the modification of the oscillation modes in the low frequency limit due to crustal plasticity can be estimated using the Newton--Raphson method with Eq.~(\ref{eq:CoupledHallModeLowFreq}) as the initial guess, giving
\begin{equation}
\omega\approx\frac{\omega_{\text{H}}\sqrt{1-\alpha}}{\sqrt{\left(1+\omega^2_{\text{A}}/\omega_s^2\right)\left(1+\omega_{\text{A}}^2(1-\alpha)/\omega_s^2\right)}}-i\frac{\check{\mu}\left[2\left(1+\omega_{\text{A}}^2/\omega_s^2\right)-\alpha\left(1+2\omega_{\text{A}}^2/\omega_s^2\right)\right]}{2\overline{\eta}\left(1+\omega_{\text{A}}^2(1-\alpha)/\omega_s^2\right)\left(1+\omega_{\text{A}}^2/\omega_s^2\right)},
\end{equation}
where $\omega_s^2$ is defined as in Eq.~(\ref{eq:AlfvenAndShearFrequency}) in terms of $\check{\mu}$. This retains the instability for $\alpha>1$, but it can be completely damped out by the plastic viscosity, with lower $\check{\mu}/\overline{\eta}$ providing greater damping. If $\omega_{\text{A}}^2/\omega_s^2\approx0.1\sigma_{\text{M}}/\sigma_y\gg 1$, which is the elasto-viscoplastic regime, then this approximates to
\begin{equation}
\omega\approx\frac{\omega_{\text{H}}\omega_s^2}{\omega_{\text{A}}^2}-i\frac{\check{\mu}}{\overline{\eta}},
\end{equation}
where the instability vanishes and the modes are only damped.

In the $|\tau\omega|\ll 1$ case, we replace $\omega_s^2\rightarrow-i\omega k^2\overline{\eta}/\rho=-i\omega k^2\nu$ in Eq.~(\ref{eq:CoupledHallModeFull}). In the high frequency limit $|\omega|^2\gg\omega_{\text{H}}^2$, this gives two modes with dispersions
\begin{align}
\omega_1=-i\frac{\nu k^2}{2}\pm\omega_{\text{A}}\sqrt{1-\frac{\nu^2k^4}{4\omega^2_{\text{A}}}}\approx -i\frac{\nu k^2}{2}\pm\omega_{\text{A}}, &&
\omega_2=-i\frac{\nu k^2}{2}\pm\omega_{\text{A}}\sqrt{1-\alpha-\frac{\nu^2k^4}{4\omega^2_{\text{A}}}}\approx -i\frac{\nu k^2}{2}\pm\omega_{\text{A}}\sqrt{1-\alpha}.
\end{align}
Since $\nu k^2\ll \check{\mu}k^2/\rho$ by assumption and $\check{\mu}k^2/\rho\ll \omega_{\text{A}}^2$ in the elasto-viscoplastic regime, the imaginary damping part of these modes is much smaller in magnitude than the real part and the modes are only slightly affected by the plastic viscosity. In the low frequency limit $|\omega|^2,\omega_{\text{H}}^2\ll\omega_{\text{A}}^2$, we obtain
\begin{equation}
\omega\approx-i\frac{\nu k^2\omega_{\text{A}}^2(2-\alpha)}{2(\omega_{\text{H}}^2(1-\alpha)+\nu^2k^4)}\pm\frac{\omega_{\text{A}}^2\omega_{\text{H}}(1-\alpha)}{\omega_{\text{H}}^2(1-\alpha)+\nu^2k^4}\sqrt{1-\frac{\alpha^2\nu^2k^4}{4\omega_{\text{H}}^2(1-\alpha)^2}}.
\end{equation}
This can be simplified if $\omega_{\text{H}}\ll\nu k^2$ to give
\begin{equation}
\omega\approx-i\frac{\nu k^2(2-\alpha)}{2(1-\alpha)}\frac{\omega_{\text{A}}^2}{\omega_{\text{H}}^2}\pm\frac{\omega_{\text{A}}^2}{\omega_{\text{H}}},
\end{equation}
which is strongly damped since $\omega_{\text{A}}\gg\omega_{\text{H}}$. But these are not the usual Hall modes, which therefore do not exist in the $|\tau\omega|\ll 1$ elasto-viscoplastic regime.

\section{Nonzero temperatures and the instability parameter space}
\label{sec:NonzeroTUnstablePS}

It was shown in Paper I that finite temperatures can suppress the instability associated with Landau quantized fermions in ideal MHD using the full Fermi--Dirac integrals for the thermodynamic quantities relevant for the instabilities i.e. the analogs to $M$, $\chi_n$ and $\mathcal{M}_n$ as defined in this paper. In this section we first describe approximations for the finite temperature corrections to $M$ and $\chi_n$, then use these approximations to determine the critical temperature for onset of the strong-field Hall MHD instability as discussed in the previous section. In doing so, we also derive an approximate criterion for the maximum number of occupied Landau levels at zero temperature $n_{\text{max}}$ required for the instability. This allows us to constrain the regions in $B$--$n_{\text{e}}$--$T$ parameter space in which the instability could be active. In this section we work in units $\hbar=c=k_B=1$.

\subsection{Low temperature corrections to thermodynamic quantities}
\label{sec:LowTCorrections}

The standard method to determine the temperature-dependence of a Fermi gas, for low to moderate temperatures, is to employ the Sommerfeld expansion (e.g.,~\citet{Ashcroft1976}). However, this expansion breaks down for the thermodynamic potentials for Landau-quantized fermions. Consider $M$: the finite temperature integral is given by
\begin{equation}
M=\left(\frac{\partial P_{\text{e}}}{\partial B}\right)_{\mu_{\text{e}},T}=-\frac{e}{2\pi^2}\sum_{n=0}^{\infty}g_n\int_{m_n}^{\infty}\frac{dE(E^2-m_n^2-neB)}{\sqrt{E^2-m_n^2}\left\{\exp[\beta(E-\mu_{\text{e}})]+1\right\}},
\label{eq:uBFiniteT}
\end{equation}
where $m_n\equiv \sqrt{m_{\text{e}}^2+2eBn}$, $\beta=T^{-1}$ and $g_n=2-\delta_{n,0}$. This can be expanded at finite temperature as
\begin{equation}
M(T)=M(T=0)+\frac{e\mu_{\text{e}}T^2}{12}\sum_{n=0}^{\nmax-1}g_n\frac{\mu_{\text{e}}^2-m_n^2+eBn}{(\mu_{\text{e}}^2-m_n^2)^{3/2}}+M_{n\geq\nmax}(T\neq0),
\end{equation}
where the terms on the right are the $T=0$ solution, the standard lowest-order Sommerfeld expansion for $n<\nmax$, and additional corrections for $n\geq\nmax$. The Sommerfeld expansion is valid for $\mu_{\text{e}}-m_n\gg T$, which is possible for $n<\nmax$, but it fails for $|\mu_{\text{e}}-m_n|\lesssim T$ or $m_n-\mu_{\text{e}}\gg T$ which occurs for $n\geq\nmax$. Moreover, while $M(0)$ is finite, its derivatives with respect to $\mu_{\text{e}}$ and $B$ are not, with divergences arising from the $n=\nmax$ term. Hence a different approximation must be used to include finite temperature corrections for $n\gtrsim n_{\text{max}}$ i.e., for small $|\mu_{\text{e}}-m_n|/T$.

After a change of integration variable to $y=\sqrt{\beta(E-m_n)}$, Eq.~(\ref{eq:uBFiniteT}) can be rewritten in the form
\begin{equation}
M=\frac{e\sqrt{T}}{2\pi^2}\sum_{n=0}^\infty g_n\sqrt{2m_n}\left[2T\int_0^{\infty}\frac{dy y^2 \sqrt{1+y^2T/(2m_n)}}{\exp(y^2-Y_n)+1}-\frac{neB}{{m_n}}\int_0^\infty\frac{dy}{\sqrt{1+y^2T/(2m_n)}[\exp(y^2-Y_n)+1]}\right],
\label{eq:uBFiniteTVariableChange}
\end{equation}
where $Y_n\equiv\beta(\mu_{\text{e}}-m_n)$. For low temperatures $\mu_{\text{e}}\gg T$ and $n<n_{\text{max}}$, $Y_n\gg 0$. This is not necessarily true for $n\geq n_{\text{max}}$. For $n=n_{\text{max}}$, the contributions to both integrals for low temperatures are exponentially suppressed for $y^2>Y_{n_{\text{max}}}$, and since $\mu_{\text{e}}\sim m_{n_{\text{max}}}$,  $Y_{n_{\text{max}}}T/(2m_{n_{\text{max}}})\ll 1$ when $y^2\lesssim Y_{n_{\text{max}}}$. So Taylor expanding to zeroth order the integrands in Eq.~(\ref{eq:uBFiniteTVariableChange}) about $Y_nT/(2m_n)\ll 1$ gives the temperature-dependent contribution to the sum over Landau levels from $n=n_{\text{max}}$
\begin{equation}
\left(M\right)_{n=n_{\text{max}}}\approx\frac{e}{\pi^2}\sqrt{\frac{m_{n_{\text{max}}}T}{2}}g_{n_{\text{max}}}\left[2T\int_0^{\infty}\frac{dy y^2}{\exp(y^2-Y_{n_{\text{max}}})+1}-\frac{n_{\text{max}}eB}{m_{n_{\text{max}}}}\int_0^\infty\frac{dy}{\exp(y^2-Y_{n_{\text{max}}})+1}\right].
\label{eq:uBFiniteTnmax}
\end{equation}
Since both integrals are $\sim 1$ for small value of $Y_n$, in this limit the second term dominates if $T\ll n_{\text{max}}eB/(2m_{n_{\text{max}}})$. Note that the two integrals only depend on a single parameter $Y_{n_{\text{max}}}$. In general, this approximation could be made more accurate by including higher-order terms in the Taylor expansions of $(1+y^2T/(2m_n))^{\pm 1/2}$.

Recall that $\nmax$ is the integer part of $p_F^2/2eB$ for Fermi momentum $p_F=\sqrt{\mu^2_{\text{e}}-m^2_{\text{e}}}$; therefore $\mu_{\text{e}}^2=p_F^2+m_{\text{e}}^2>2eB\nmax+m_{\text{e}}^2=m_{\nmax}^2$. At $T=0$ there are no terms in the sums with $n>\nmax$, but at nonzero $T$ these terms are present. Hence there are \textit{two} cases to consider.

First, consider what happens when $\mu_{\text{e}}$ is near but just above $m_{\nmax}$. In this first case, all terms with $n<\nmax$ can be approximated
by their $T=0$ values, assuming that 
\begin{equation}
\mu_{\text{e}}-m_{\nmax-1}\simeq m_{\nmax}-m_{\nmax-1}=m_{\nmax}\sqrt{m_{\nmax}^2-2eB}\simeq\frac{eB}{m_{\nmax}}\simeq\frac{eB}{\mu_{\text{e}}}\gg T,
\end{equation}
which is equivalent to the requirement that the Landau level spacing at the Fermi surface is larger than the temperature. We are therefore
left with considering just the terms $n\geq\nmax$. The dominant terms are
\begin{equation}
-\frac{n e^2B}{\pi^2}\sqrt{\frac{2T}{m_n}}\int_0^\infty\frac{\text{d}y}{\sqrt{(1+y^2T/2m_n)}[\exp(y^2-Y_n)+1]}
\simeq -\frac{n e^2B}{\pi^2}\sqrt{\frac{2T}{m_n}}\int_0^\infty\frac{\text{d}y}{\exp(y^2-Y_n)+1}
\equiv -\frac{n e^2BF_{-1/2}(Y_n)}{2\pi^{3/2}}\sqrt{\frac{2T}{m_n}}~,
\label{eq:IzeroYnu}
\end{equation}
where $F_j(Y)$ is the complete Fermi--Dirac integral. Since $Y_{\nmax}>0$ in this case, the integrand is largest near $y=\sqrt{Y_{\nmax}}$, and cuts off exponentially within $\delta y\sim 1$ of its peak value. As a result $I_0(Y_{\nmax})\sim 1$ for small values of $Y_{\nmax}$; the contribution from $n=\nmax$ to the magnetization is of order $(ep_F^2/\pi^2)\sqrt{T/\mu_{\text{e}}}$, which goes to zero as $T\to 0$. The $T=0$ limit is approached once $Y_{\nmax}$ is larger than $\sim$ a few i.e., for $\mu_{\text{e}}-m_{\nmax}\gtrsim ({\rm a\, few})\times T$, so $\mu_{\text{e}}$ is still very near $m_{\nmax}\gg T$. The contribution from $n=\nmax+1$ is suppressed because $Y_{\nmax+1}=(\mu_{\text{e}}-m_{\nmax+1})/T\simeq (m_{\nmax}-m_{\nmax+1})/T\simeq -eB/\mu_{\text{e}} T$, which is very negative for $T$ small compared to the Landau level spacing, and therefore $I_0(Y_{\nmax+1})\sim \exp(-|Y_{\nmax+1}|)\ll 1$.

Next consider what happens when $\mu$ is near but just below $m_{\nmax}$. In this case, the contributions from $n<\nmax$ are once again close to their $T=0$ values. The contribution from $n=\nmax$ can still be found from Eq. (\ref{eq:IzeroYnu}) but now $Y_{\nmax}<0$. Although this contribution is exponentially small for large $|Y_{\nmax}|$, for $0>\mu_{\text{e}}-m_{\nmax}\gtrsim -T$  it is of the same order as the contribution for $0<\mu_{\text{e}}-m_{\nmax}\lesssim T$. Thus, the magnetization is approximately
\begin{equation}
M\simeq M_{0\leq n < \nmax}(B,\mu_{\text{e}},0)-\frac{ep_F^2F_{-1/2}(Y_{\nmax})}{4\pi^{3/2}}\sqrt{\frac{2T}{\mu_{\text{e}}}}~
\label{eq:MApprox}
\end{equation}
where the first term on the right-hand side is the contribution from all Landau levels below $n=\nmax$ calculated at $T=0$, and where $Y_{\nmax}$ can be positive or negative. Completely analogous reasoning for the electron density
\begin{equation}
n_{\text{e}}=\frac{eB}{2\pi^2}\sum_{n=0}^\infty g_n\int_{m_n}^\infty\frac{\text{d}E\,E}{\sqrt{E^2-m_n^2}\,\{\exp[(E-\mu_{\text{e}})/T]+1\}}
=\frac{eB\sqrt{T}}{\pi^2\sqrt{2}}\sum_{n=0}^\infty{g_n\sqrt{m_n}}\int_0^\infty\frac{dy(1+y^2T/m_n)}{\sqrt{1+y^2T/(2m_n)}[\exp(y^2-Y_n)+1]}
\end{equation}
leads to the approximation
\begin{equation}
n_{\text{e}}\simeq n_{\text{e},0\leq n < \nmax}(B,\mu_{\text{e}},0)+\frac{eB\sqrt{2T\mu_{\text{e}}}F_{-1/2}(Y_{\nmax})}{2\pi^{3/2}}.
\label{eq:nApprox}
\end{equation}
In the $n=\nmax$ terms in Eqs. (\ref{eq:MApprox}) and (\ref{eq:nApprox}) only the leading terms in $T$ have been retained and $\mu_{\text{e}}$ has been substituted for $m_{\nmax}$ since the most temperature-sensitive dependence in the derivatives $\partial (M,n_e)/\partial B$ will arise from differentiating
$I_0(Y)$.

Since the instability criterion as found in Eq.~(\ref{eq:UncoupledHallMode}) depends on $\chi_n$, we also want to determine the lowest-order temperature-dependent corrections to this quantity. From Eq.~(\ref{eq:dM}) we have
\begin{equation}
\left(\frac{\partial H}{\partial B}\right)_{n_{\text{e}},T}=\left(\frac{\partial H}{\partial B}\right)_{\mu_{\text{e}},T}-\left(\frac{\partial H}{\partial\mu_{\text{e}}}\right)_{B,T}
\left(\frac{\partial n_{\text{e}}}{\partial\mu_{\text{e}}}\right)_{B,T}^{-1}\left(\frac{\partial n_{\text{e}}}{\partial B}\right)_{\mu_{\text{e}},T}.
\label{eq:dHdB}
\end{equation}
When $\beta(\mu_{\text{e}}-m_n)\lesssim 1$ i.e., when $(\mu_{\text{e}}^2-m_{\text{e}}^2)/(2eB)$ is nearly an integer and where the instability occurs, we can drop the contribution from the $n=n_{\text{max}}+1$ Landau level in Eq.~(\ref{eq:MApprox}--\ref{eq:nApprox}). Here we hold $n_{\text{e}}$ fixed because it, as opposed to $\mu_{\text{e}}$, is a natural variable for the internal energy density using which the equations of motion were derived. Taking the required partial derivatives for computing Eq.~(\ref{eq:dHdB}) (following e.g., Paper I Appendix B (online supplement) for the required partial derivatives of $H$ at $T=0$) gives
\begin{subequations}
\begin{align}
\left(\frac{\partial H(B,\mu_{\text{e}},T)}{\partial B}\right)_{\mu_{\text{e}},T}&\simeq\left(\frac{\partial H_{0\leq n < \nmax}(B,\mu_{\text{e}},0)}{\partial B}\right)_{\mu_{\text{e}},T}-\frac{e^2\nmax}{\sqrt{\pi}}\sqrt{\frac{2\mu_{\text{e}}}{T}}F_{-3/2}(Y_{n_{\text{max}}})
\nonumber\\
&\simeq 1-\frac{2e^2\nmax}{\pi}\Bigg[2\left\{v_F\Shalf-(1-v_F^2)\Sigma_S-2v_F^2\SSn\right\}-\left(\frac{2+v_F^2}{v_F}\right)\Shalf + v_F\Shalfm - v_F(1-v_F^2)\Sc
\nonumber\\
&\qquad\qquad\qquad\qquad + 2\left(\frac{1}{v_F^2}-1\right)\Sigma_S+\sqrt{\frac{\pi\mu_{\text{e}}}{2T}}F_{-3/2}(Y_{n_{\text{max}}})\Bigg],
\label{eq:dHdBApprox}
\\
\left(\frac{\partial H(B,\mu_{\text{e}},T)}{\partial\mu_{\text{e}}}\right)_{B,T}&\simeq\left(\frac{\partial H_{0\leq n < \nmax}(B,\mu_{\text{e}},0)}{\partial\mu_{\text{e}}}\right)_{B,T}+\frac{2e\mu_{\text{e}}}{\pi}\sqrt{\frac{\pi\mu_{\text{e}}}{2T}}F_{-3/2}(Y_{n_{\text{max}}})
=-\frac{2e\mu_{\text{e}}}{\pi}\left[3\Shalf-\Shalfm-\sqrt{\frac{\pi\mu_{\text{e}}}{2T}}F_{-3/2}(Y_{n_{\text{max}}})\right],
\\
&=-4\pi\left(\frac{\partial n_{\text{e}}(B,\mu_{\text{e}},T)}{\partial B}\right)_{\mu_{\text{e}},T}
\nonumber
\\
\left(\frac{\partial n_{\text{e}}(B,\mu_{\text{e}},T)}{\partial\mu_{\text{e}}}\right)_{B,T}&\simeq\left(\frac{\partial n_{\text{e},0\leq n < \nmax}(B,\mu_{\text{e}},0)}{\partial\mu_{\text{e}}}\right)_{B,T}+\frac{\mu_{\text{e}}^2}{2\pi^2\nmax}\sqrt{\frac{\pi\mu_{\text{e}}}{2T}}F_{-3/2}(Y_{n_{\text{max}}})
=\frac{\mu_{\text{e}}^2}{2\pi^2\nmax}\left[\Shalfm+\sqrt{\frac{\pi\mu_{\text{e}}}{2T}}F_{-3/2}(Y_{n_{\text{max}}})\right],
\label{eq:dndmuApprox}
\end{align}
\end{subequations}
where $v_F^2\equiv (1-m_{\text{e}}^2/\mu_{\text{e}}^2)$ and we use $\mu_{\text{e}}^2\approx 2eB\nmax$ so
\begin{subequations}
\begin{align}
\Shalf={}&\sum_{n=0}^{\nmax-1}\frac{g_n}{2}\sqrt{1-\frac{n}{\nmax}}, & \Sigma_S={}&\sum_{n=0}^{\nmax-1}\frac{g_n}{2}\text{arsinh}\sqrt{\frac{v_F^2(\nmax/n-1)}{v_F^2+(1-v_F^2)\nmax/n}},
\nonumber
\\
\SSn={}&\sum_{n=1}^{\nmax-1}\frac{n}{\nmax}\text{arsinh}\sqrt{\frac{v_F^2(\nmax/n-1)}{v_F^2+(1-v_F^2)\nmax/n}}, & \Shalfm={}&\sum_{n=0}^{\nmax-1}\frac{g_n}{2\sqrt{1-n/\nmax}},
\label{eq:SigmaExpressions}
\\
\Sc={}&\sum_{n=0}^{\nmax-1}\frac{g_n}{2}\frac{n\sqrt{1-n/\nmax}}{\nmax(1+v_F^2(n/\nmax-1))}, & \Sigma_{-3/2}={}&\sum_{n=0}^{\nmax-1}\frac{g_n}{2(1-n/\nmax)^{3/2}}.
\nonumber
\end{align}
\end{subequations}
The complete Fermi-Dirac integral $F_{-3/2}(Y)$ has asymptotic forms $F_{-3/2}(Y)\sim 1\sqrt{\pi}/(4\sqrt{Y})$ for $Y\gg 1$ and $F_{-3/2}(Y)\sim\sqrt{\pi}\exp(Y)/2$ for $Y\ll 0$, and a peak value of $F_{-3/2}(Y\approx1.1)\approx0.4463$. 

\begin{figure}
\center
\centering
{\includegraphics[width=0.47\linewidth]{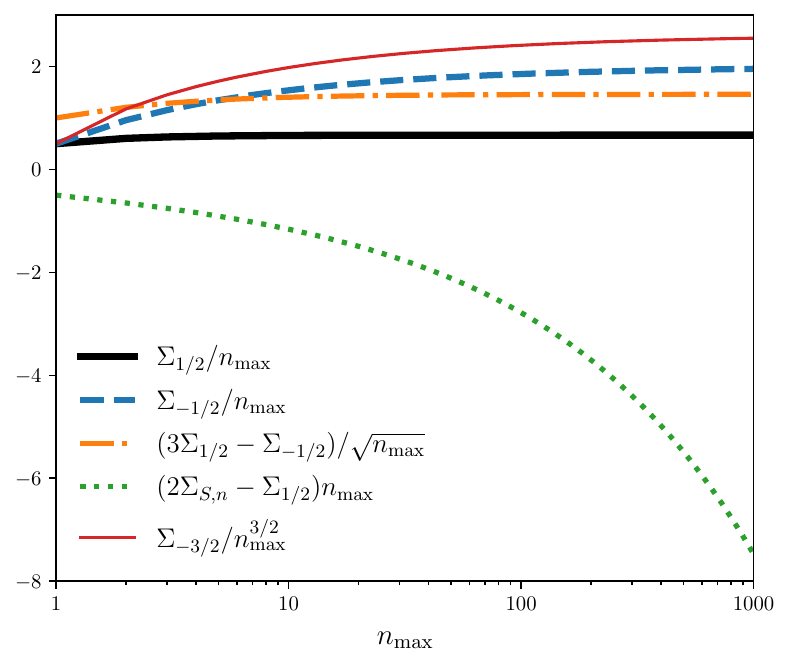}}
\caption{Scaled asymptotic forms of different sums as defined in Eq.~(\ref{eq:SigmaExpressions}) as a function of $\nmax$, justifying the scalings given in Eq.~(\ref{eq:SumScalings}). $v_F^2=1$ is assumed in computing $\SSn$.}
\label{fig:dIdYandSumsPlot}
\end{figure}

Combining Eq.~(\ref{eq:dHdBApprox}--\ref{eq:dndmuApprox}), working in the ultrarelativistic regime $v_F\approx 1$, and only retaining terms $\mathcal{O}(T^{-1/2})$ and $\mathcal{O}(T^0)$ gives the desired approximation for $(\partial H/\partial B)_{n_{\text{e}}}$
\begin{equation}
\left(\frac{\partial H}{\partial B}\right)_{n_{\text{e}}}\simeq 1-\frac{4e^2\nmax}{\pi}\left(\Shalf-2\SSn\right)+\frac{6e^2\nmax\Shalf}{\pi}\left[-1+\frac{3\Shalf}{\Shalfm+\sqrt{\pi\mu_{\text{e}}/(2T)}F_{-3/2}(Y_{n_{\text{max}}})}\right].
\label{eq:dHdBFiniteTApprox}
\end{equation}

\subsection{Unstable region of $B$--$n_{\text{e}}$--$T$ parameter space}

\begin{figure}
\center
\centering
\subfloat[]{\includegraphics[width=0.45\linewidth]{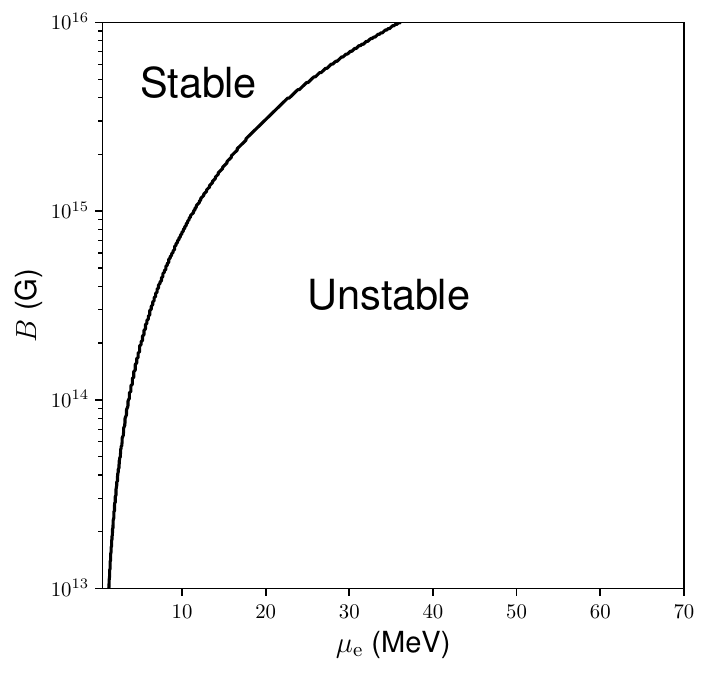}}
  \hfill
  \subfloat[]{\includegraphics[width=0.45\linewidth]{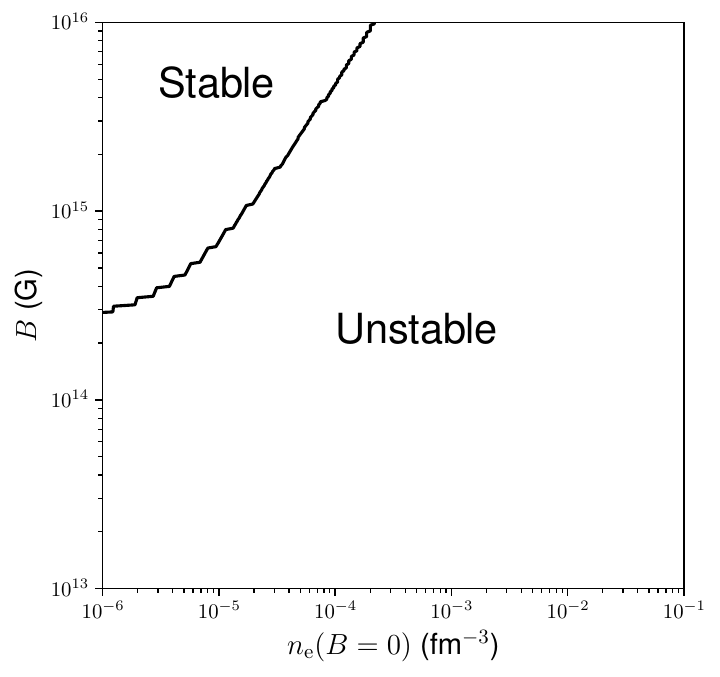}}
\caption[]{The contour of constant $\nmax=\sqrt{\pi/(4e^2)}$ (approximating $\nmax$ as a real number and not an integer) as a function of $B$ and $\mu_{\text{e}}$ (a) and $B$ and $n_{\text{e}}(T=0)$ (b), with the potentially stable and unstable regions of parameter space split by these contours labelled.}
\label{fig:TZeroInstabilityParamSpace}
\end{figure}

Since the instability criterion is roughly $1-4\pi\chi_n<0$ or equivalently $\partial H/\partial B|_{n_{\text{e}}}<0$, we can use Eq.~(\ref{eq:dHdBFiniteTApprox}) to estimate the region of parameter space where the strong field Hall MHD instability may be active. We can first estimate the values of the sums in Eq.~(\ref{eq:SigmaExpressions}). These expressions are shown in Figure~\ref{fig:dIdYandSumsPlot} panel (b) as a function of $\nmax$, which shows that they have the following asymptotic forms for large $\nmax$ in the $m_{\text{e}}\rightarrow0$ limit:
\begin{align}
\Shalf\simeq 2\nmax/3, && \Shalfm\simeq 2\nmax, && 3\Shalf-\Shalfm\simeq 1.5\sqrt{\nmax}, && \Sigma_{-3/2}\simeq 2.6\nmax^{3/2}, && \Shalf-2\SSn\sim 1/\nmax.
\label{eq:SumScalings}
\end{align}
In the zero-temperature limit and working to leading order in $\nmax$, Eq.~(\ref{eq:dHdBFiniteTApprox}) reduces to
\begin{equation}
\left(\frac{\partial H}{\partial B}\right)_{n_{\text{e}}}\simeq 1-\frac{4e^2\nmax}{\pi}\left(\Shalf-2\SSn\right)-\frac{6e^2\nmax\Shalf}{\pi}\approx 1-\frac{4e^2\nmax^2}{\pi}.
\label{eq:dHdBZeroTApprox}
\end{equation}
This condition implies that $n_{\text{max}}\gtrsim\sqrt{\pi/(4e^2)}=10.37$ is necessary for the instability (assuming $n_{\text{e}}$ is held fixed in the derivatives), and since $\nmax$ is determined by $B$ and $\mu_{\text{e}}$, it sets the possible region in parameter space where the instability could be active under the given approximations. Figure~\ref{fig:TZeroInstabilityParamSpace} shows the potentially unstable region in $B$--$\mu_{\text{e}}$ and $B$--$n_{\text{e}}$ parameter spaces at $T=0$ and in the large $\nmax$ approximation. Since $n_{\text{e}}$ is a function of $B$, in the latter we use the $n_{\text{e}}(B=0)$ value as an independent variable.

\begin{figure}
\center
\centering
\subfloat[]{\includegraphics[width=0.497\linewidth]{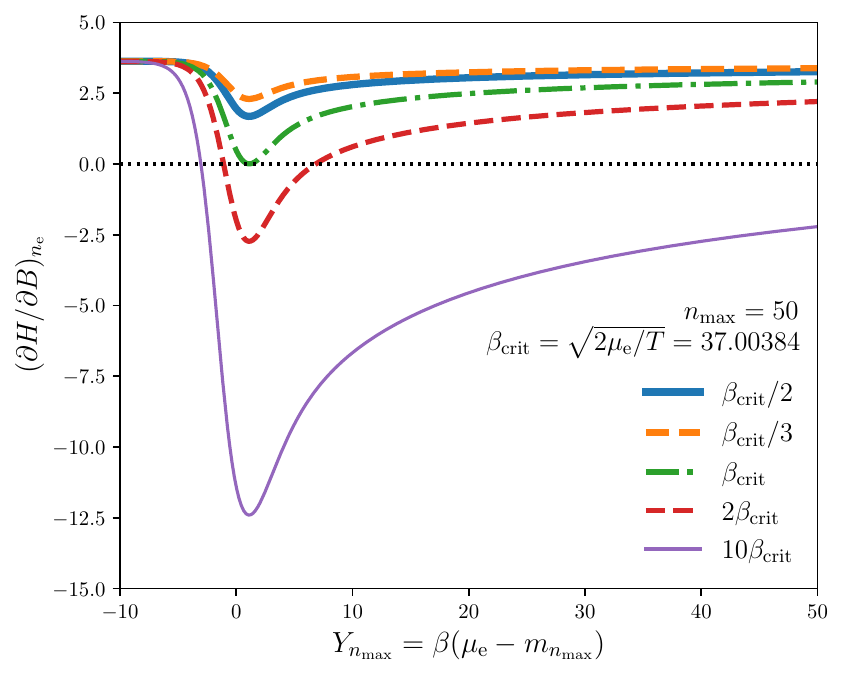}}
  \hfill
  \subfloat[]{\includegraphics[width=0.503\linewidth]{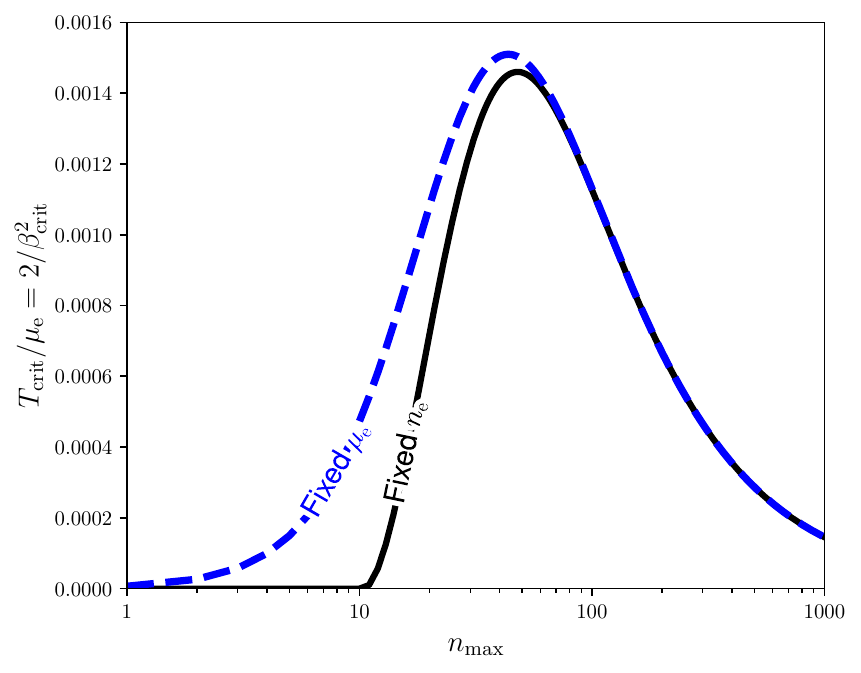}}
\caption[]{(a) Eq.~(\ref{eq:dHdBFiniteTApprox}) for fixed $\nmax=50$ as a function of $Y$. Points above $(\partial H/\partial B)_{n_{\text{e}}}=0$ are stable. $\beta_{\text{crit}}=37.00384$, defined by Eq.~(\ref{eq:betacrit}), is the value marking the transition between stability and instability, and Eq.~(\ref{eq:dHdBFiniteTApprox}) is plotted for this value and for different multiples of it. $v_F^2=1$ is assumed. (b) $T_{\text{crit}}/\mu_{\text{e}}$ as a function of $\nmax$ computed using Eq.~(\ref{eq:betacrit}) for $n_{\text{e}}$ fixed and Eq.~(\ref{eq:dHdBApprox}) for $\mu_{\text{e}}$ fixed. $v_F^2=1$ is assumed for both curves. Note the critical temperature is zero for $n_{\text{max}}\leq 10$ only in the fixed $n_{\text{e}}$ case.}
\label{fig:dHdBandTcritvsnmax}
\end{figure}

Including the lowest-order finite temperature corrections, we find that the condition for instability is
\begin{equation}
\frac{3\Shalf}{\Shalfm+\sqrt{2\mu_{\text{e}}/T}I'_0(Y_{n_{\text{max}}})}<1-\frac{\pi}{6e^2\nmax\Shalf}-\frac{2(2\SSn-\Shalf)}{3\Shalf}=\frac{5}{3}-\frac{4\SSn}{3\Shalf}
-\frac{\pi}{6e^2\nmax\Shalf}.
\label{eq:betacritDerivation}
\end{equation}
For values of $\nmax\lesssim 10$, the right-hand side of this equation cannot be positive, and since left-hand side of the equation is always positive the instability is not active for such low values of $\nmax$. Assuming that $\nmax$ is large enough to make the right-hand side positive, and noting that $I'_0(Y)\lesssim 0.3955$, the necessary condition for instability is
\begin{equation}
\sqrt{\frac{2\mu_{\text{e}}}{T}}>2.528\left(\frac{3\Shalf}{5/3-4\SSn/(3\Shalf)-\pi/(6e^2\nmax\Shalf)}-\Shalfm\right)\equiv\beta_{\text{crit}}.
\label{eq:betacrit}
\end{equation}
and hence a critical temperature above which the instability is inactive $T_{\text{crit}}=2\mu_{\text{e}}/\beta_{\text{crit}}^2$. For large $n_{\text{max}}$ Eq.~(\ref{eq:betacrit}) reduces to $\beta_{\text{crit}}\approx 2.528(3\Shalf-\Shalfm)\approx 3.8\sqrt{n_{\text{max}}}$. This implies a critical temperature of
\begin{equation}
T_{\text{crit}}\approx 1.6\times10^8\left(\frac{\mu_{\text{e}}}{10\text{ MeV}}\right)\left(\frac{n_{\text{max}}}{100}\right)^{-1}\text{ K}.
\label{eq:TcritLargenmax}
\end{equation}
Figure~\ref{fig:dHdBandTcritvsnmax} panel (a) shows Eq.~(\ref{eq:dHdBFiniteTApprox}) as a function of $Y$ for fixed $\nmax$ and different values of $\sqrt{2\mu_{\text{e}}/T}$, labelled by their scaling compared to $\beta_{\text{crit}}=\sqrt{2\mu_{\text{e}}/T_{\text{crit}}}$ as defined by Eq.~(\ref{eq:betacrit}). It demonstrates the suppression of the instability as the temperature is increased, with the range of unstable values of $Y$ decreasing to a single point at $T_{\text{crit}}$.

\begin{figure}
\center
\centering
\subfloat[]{\includegraphics[width=0.5\linewidth]{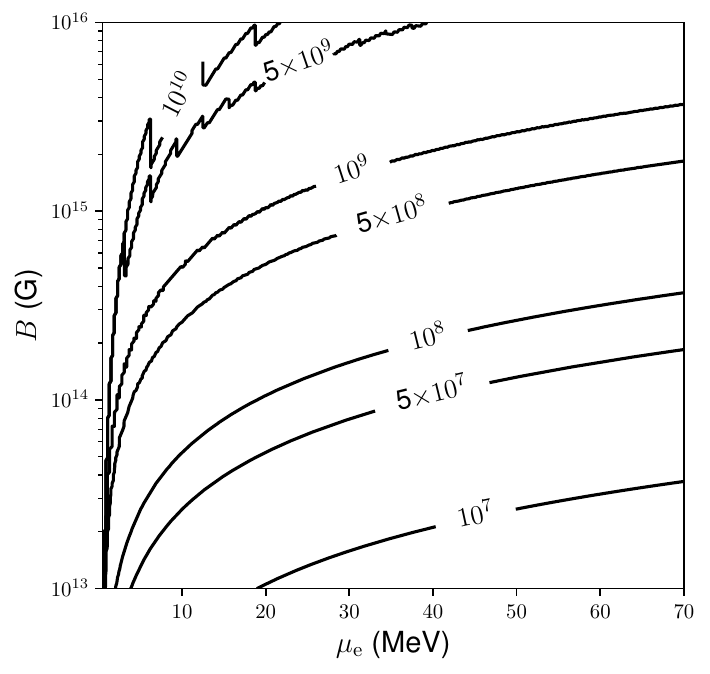}}
  \hfill
  \subfloat[]{\includegraphics[width=0.5\linewidth]{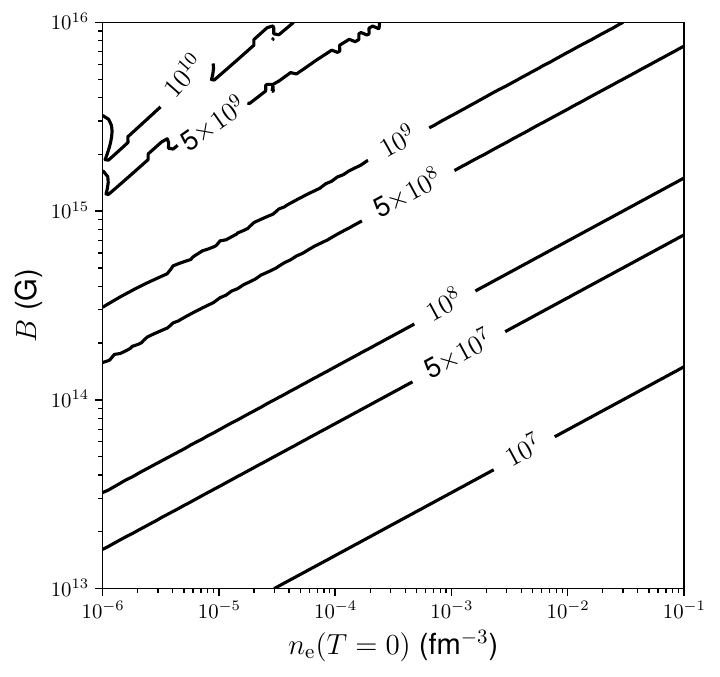}}
\caption[]{Critical temperature $T_{\text{crit}}$ contours in K as a function of $B$ and $\mu_{\text{e}}$ (left) and $B$ and $n_{\text{e}}(T=0)$ (right), computed using the large $\nmax$ approximation for $T_{\text{crit}}$ (Eq.~(\ref{eq:TcritLargenmax})).}
\label{fig:CriticalTContoursLargenmax}
\end{figure}

The contours of Figure~\ref{fig:CriticalTContoursLargenmax} show the critical temperature for the instability in $B$--$\mu_{\text{e}}$ and $B$--$n_{\text{e}}(T=0)$ parameter space using the large $\nmax$ approximation Eq.~(\ref{eq:TcritLargenmax}). The critical temperature increases as the density decreases and the magnetic field increases i.e. as $n_{\text{max}}$ increases. But the approximation is not valid in the upper left-hand regions of each plot. 

\begin{figure}
\center
\centering
\subfloat[]{\includegraphics[width=0.5\linewidth]{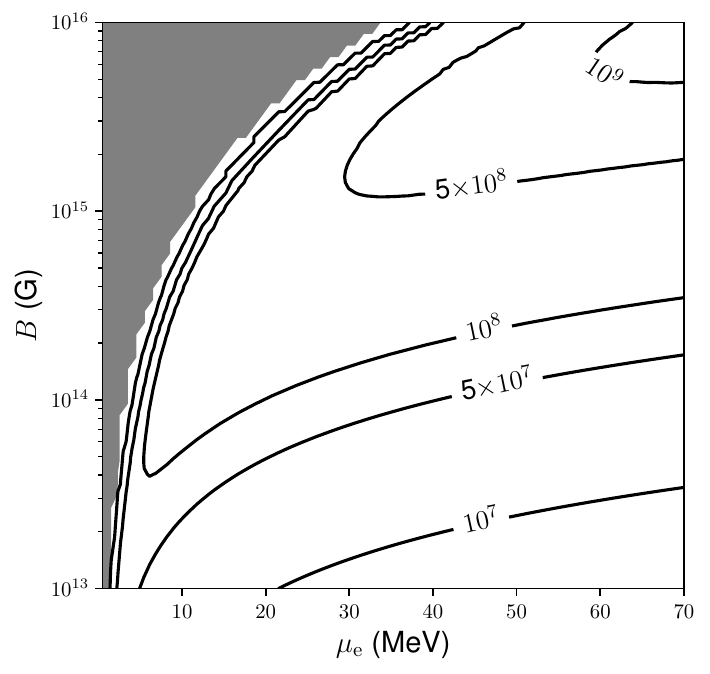}}
  \hfill
  \subfloat[]{\includegraphics[width=0.5\linewidth]{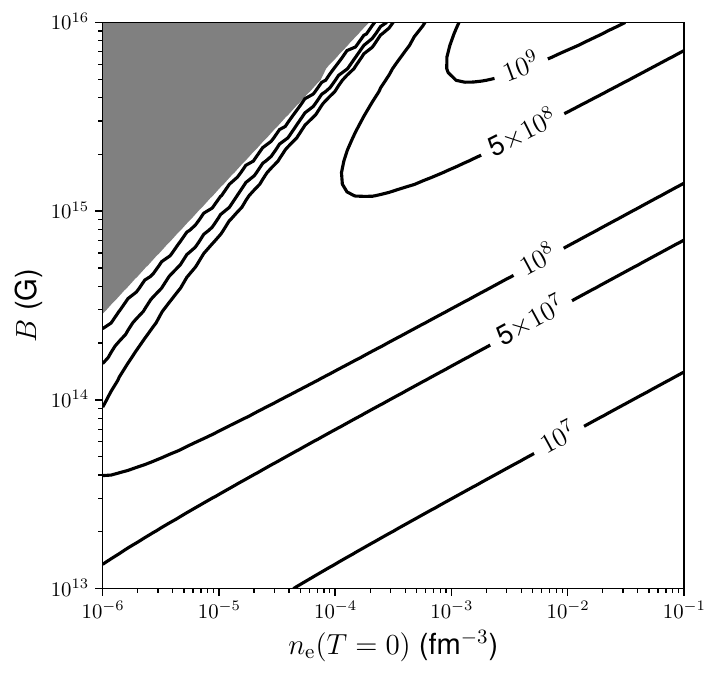}}
\caption[]{Critical temperature $T_{\text{crit}}$ contours in K as a function of $B$ and $\mu_{\text{e}}$ (a) and $B$ and $n_{\text{e}}(T=0)$ (b), computed using $T_{\text{crit}}/\mu_{\text{e}}=2/\beta_{\text{crit}}^2$ and the full expression for $\beta_{\text{crit}}$, Eq.~(\ref{eq:betacrit}). The grey region is that for which the right-hand side of Eq.~(\ref{eq:betacritDerivation}) is negative, which is always stable.}
\label{fig:CriticalTContours}
\end{figure}

Figure~\ref{fig:CriticalTContours} shows the contours of fixed critical temperature for the instability using the full form of $\beta_{\text{crit}}$ in Eq.~(\ref{eq:betacrit}). The critical temperature generally increases as the density decreases and the magnetic field increases i.e., as $n_{\text{max}}$ increases.  Moving from high $\mu_{\text{e}}$ to low $\mu_{\text{e}}$ along nearly constant $B$, the contours eventually turn around and follow curves of nearly constant $\mu_{\text{e}}$ for increasing $B$. This behaviour can be understood by plotting $T_{\text{crit}}$ as a function of $\nmax$, as is done in Figure~\ref{fig:dHdBandTcritvsnmax} (b). This shows that the critical temperature is zero for $\nmax\leq 10$, increases to a peak value for $\nmax=48$, where  $T_{\text{crit}}/\mu_{\text{e}}\approx0.00146$, and then decreases for larger $\nmax$. This explains the ``turn-around'' in the critical temperature contours: on a curve of fixed $\mu_{\text{e}}$ in the left panel of Figure~\ref{fig:CriticalTContours}, as $B$ is increased, the curve can intersect the same constant $T_{\text{crit}}$ contour twice. As $B$ is increased, $\nmax$ decreases, and so going from right to left across Figure~\ref{fig:dHdBandTcritvsnmax} panel (b), $T_{\text{crit}}$ will have the same value for two different values of $\nmax$ or two different values of $B$ for fixed $\mu_{\text{e}}$. $T_{\text{crit}}$ at fixed $\mu_{\text{e}}$ is also plotted for comparison: this is found by using Eq.~(\ref{eq:dHdBApprox}) while setting $\partial H/\partial B|_{\mu_{\text{e}},T}=0$ and taking $I'_0(Y)=0.3955$. Note that there is no critical value of $n_{\text{max}}$ below which the instability is absent, unlike in the fixed $n_{\text{e}}$ case. For large $n_{\text{max}}$ the two curves are identical.

Figure~\ref{fig:dHdBandTcritvsnmax} panel (b) also informs us of how the system evolves under the instability. Holding $\mu_{\text{e}}$ fixed, for a system at $T_{\text{crit}}$ and if no heating occurs, then the evolution of $\nmax$ will occur along a constant $T_{\text{crit}}/\mu_{\text{e}}$ line from higher to lower $\nmax$ i.e., the field will increase until reaching the branch of $T_{\text{crit}}/\mu_{\text{e}}$ left of the peak at $\nmax=48$. An approximation for the amount the field could grow by the instability is found by taking the two values of $\nmax$ at which the $T_{\text{crit}}/\mu_{\text{e}}$ line intersects the curve and using $B\approx(\mu^2_{\text{e}}-m^2_{\text{e}})/(2e\nmax)$. So for fixed $\mu_{\text{e}}=25$ MeV and $T_{\text{crit}}/\mu_{\text{e}}=0.001$ i.e., $T_{\text{crit}}=2.9\times10^8$ K, as the $T_{\text{crit}}/\mu_{\text{e}}=0.001$ line intersects the curve at $\nmax=24$ and $\nmax=119$, as the field evolves isothermally it increases from $4.4\times10^{14}$ G to $2.2\times10^{15}$ G, a factor of five growth. This growth will not occur continuously, as the fluid stabilizes itself when not near a filled Landau level region of parameter space, so this is an estimate of the maximum growth that could occur for a particular region. This is of course a simplification, since (1) the instability criterion is more complicated than assumed in this section, and also depends on the relative strengths of components of the magnetic field; (2) the evolution will not happen isothermally, and in fact the field growth by the instability could lead to heating as we discuss in the next section.
 
\begin{figure}
\center
\center
\centering
\subfloat[]{\includegraphics[width=0.5\linewidth]{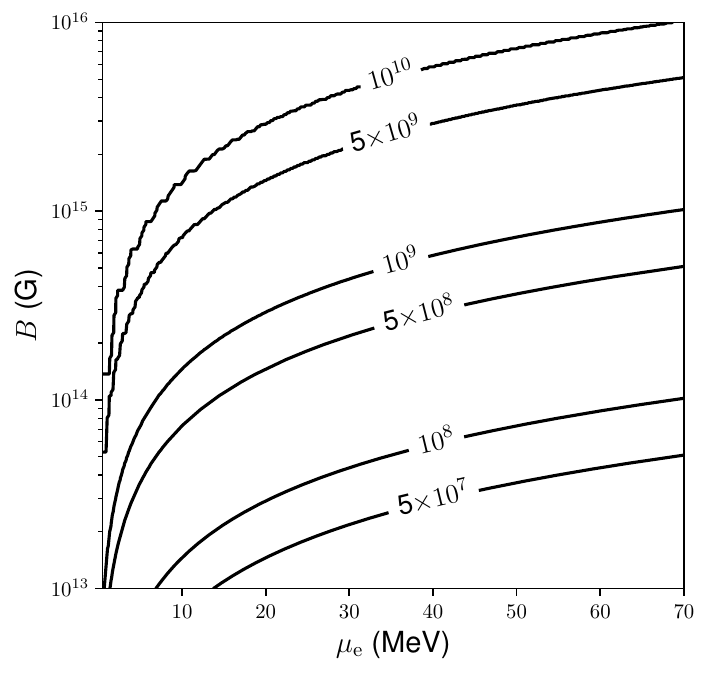}}
  \hfill
  \subfloat[]{\includegraphics[width=0.5\linewidth]{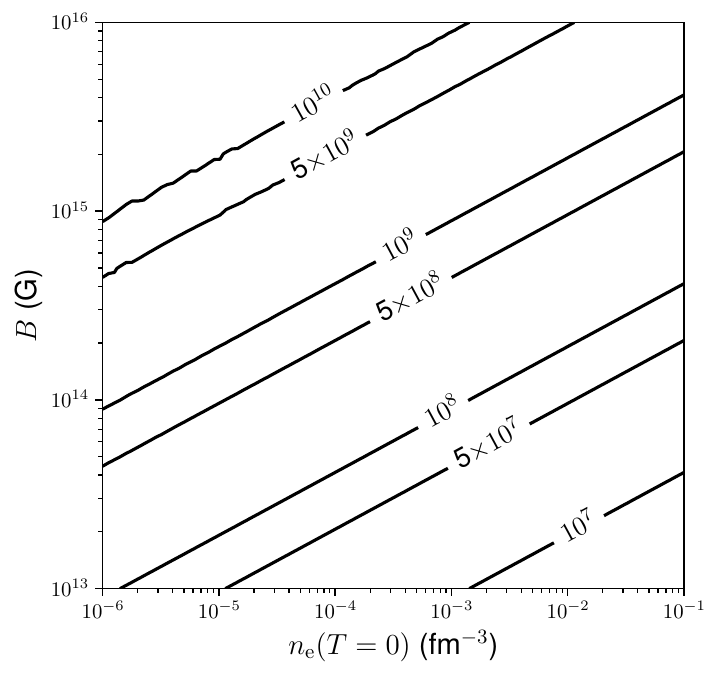}}
\caption[]{Contours of constant $T_B$, the temperature at which the spacing between adjacent Landau levels equals the thermal energy and an indicator of thermal smearing of the Landau levels. Values are in units of K. Plotted as a function of $B$ and $\mu_{\text{e}}$ (a) and $B$ and $n_{\text{e}}(T=0)$ (b).}
\label{fig:T_B}
\end{figure}

The assumption that the electrons will be Landau quantized breaks down when the energy spacing between successive Landau levels is near the thermal energy, at which point the Landau levels are smeared out. The temperature at which this occurs can be estimated using~\citep{Harding2006}
\begin{equation}
T_B=\left(\sqrt{m_{\text{e}}^2+2eB(\nmax+1)}-\sqrt{m_{\text{e}}^2+2eB\nmax}\right).
\end{equation}
The contours of constant $T_B$ are plotted in Figure~\ref{fig:T_B} as a function of $B$ and $\mu_{\text{e}}$ or $n_{\text{e}}(T=0)$. In general, at given location in parameter space, $T_B$ will be about a factor of five larger than $T_{\text{crit}}$, and hence the instability will always be inactive before significant thermal smearing of the Landau levels occurs. As temperature is increased we need to include higher Landau levels than $n=n_{\text{max}}+1$ in our computation of $T_{\text{crit}}$, but the temperatures at which this is necessary are above $T_B$, so since $T_B>T_{\text{crit}}$ the system will already be stable at these temperatures.

We note that the results here are limited by the approximations used, since Section~\ref{sec:LowTCorrections} assumed low temperatures $T\leq T_B$. Overall the most important number for determining whether the instability can be active in a certain region of $B$--$\mu_{\text{e}}$ or $B$--$n_{\text{e}}(T=0)$ parameter space is $\nmax$, and it must be greater than around $10$ for the instability assuming the required partial derivatives are computed at fixed $n_{\text{e}}$. Thus strongly quantizing fields $\nmax\approx 0$ will not be unstable. The most interesting range of $\nmax$ for the instability is between 10 and $\sim 100$, as $T_{\text{crit}}$ drops below the expected temperatures for a magnetar crust as $\nmax$ increases too far beyond this point. Our analysis also applies to the core and Paper I, since the instability condition there also depends on the negativity of $1-4\pi\chi_n$. The range in parameter space that was examined here was chosen for its relevance to the crust.

\section{Physical implications}
\label{sec:PhysicalImplications}

\subsection{Enhanced Ohmic dissipation}
\label{sec:OhmicDiss}

Since the electric current density is related by Amp\`{e}re's Law to the curl of $H$, not the curl of $B$, the Ohmic dissipation will be modified by the same terms which lead to an instability. Estimating the current density as
\begin{equation}
J^i_{e}=\frac{c}{4\pi}\epsilon^{ijk}\nabla_jH_k=\frac{c}{4\pi}\left[\frac{H}{B}\epsilon^{ijk}\nabla_jB_k-4\pi\left(\chi_n-\frac{M}{B}\right)\epsilon^{ijk}\nabla_jB\hat{B}_k\right],
\end{equation}
the Ohmic heating rate per unit volume is given by
\begin{equation}
\dot{q}_{\text{O}}=\frac{1}{\sigma}J^2_e\approx\frac{c^2}{16\pi^2\sigma}\left[\epsilon^{ijk}\nabla_jB_k-4\pi\chi_n\epsilon^{ijk}\nabla_jB\hat{B}_k\right|^2
\label{eq:OhmicDiss}
\end{equation}
where we used the approximations $|\chi_n|\gg |M|/B$ and $1\gg 4\pi|M|/B$ in the second line. To estimate the relative size of this heating rate compared to the $H=B$ result, assuming a straight field we obtain
\begin{equation}
\dot{q}_{\text{O}}=\frac{c^2}{16\pi^2\sigma}\left(1-4\pi\chi_n\right)^2|\epsilon_{ijk}\nabla_jB_k|^2.
\end{equation}
So the Ohmic dissipation is enhanced by a factor of $(1-4\pi\chi_n)^2$ compared to the result ignoring Landau quantization in regions where $p_F^2/(2eB)$ is close to $\nmax$. Even at high temperatures $|4\pi\chi_n|$ can still be of order one as Figure~\ref{fig:dHdBandTcritvsnmax} panel (a) shows, so the heating can be increased by a factor $\gtrsim 4$ or more. The increased Ohmic heating is possible for both signs of $\chi_n$, unlike the instability which requires it to be negative. However, since this enhancement is only present in limited spatial regions, the overall increase in Ohmic dissipation may be modest.

This estimate of increased Ohmic dissipation has made the inaccurate assumption that the conductivity is unchanged by the Landau quantization of electrons. For strongly-quantizing fields $\sigma$ can change by a factor $\sim10$ compared to its classical value due to Shubnikov--de Haas oscillations~\citep{Potekhin1999a}. However, for $\nmax$ in the unstable regime $\gtrsim 10$, $\sigma$ does not differ from its classical value by more than a factor of $\sim 2$. Thus the enhancement of the Ohmic dissipation due to $\chi_n$ may not be completely suppressed by an increase in the conductivity in those regions of $B$--$n_{\text{e}}$ parameter space where we expect an instability. Unlike the Shubnikov--de Haas oscillations and the instability, the increase in the Ohmic dissipation is likely to persist at higher temperatures to some extent.

\subsection{Domain formation and domain-formation heating}
\label{sec:DomainFormation}

As discussed in previous studies of magnetized neutron star crusts~\citep{Blandford1982,Suh2010,Wang2013,Wang2016}, regions of parameter space where $\chi_n>1/(4\pi)$ are thermodynamically unstable to the formation of magnetic domains. This Shoenberg effect~\citep{Shoenberg1984} and the resulting Condon domains~\citep{Condon1966,Egorov2010} are well studied in laboratory settings. The resulting magnetic domain-forming regions are not composed of alternating domains of opposite magnetization as in ferromagnets. This fact has sometimes been obscured in the literature by the tendency to split the grand potential $\Omega$ and hence the magnetization derived from it into ``non-oscillatory'' and ``oscillatory'' parts, the latter of which is split into domains of opposite magnetization, but when added to the former the overall magnetization of neighbouring domains will be aligned but have  different magnitudes.

\begin{figure}
\center
\includegraphics[width=0.68\linewidth]{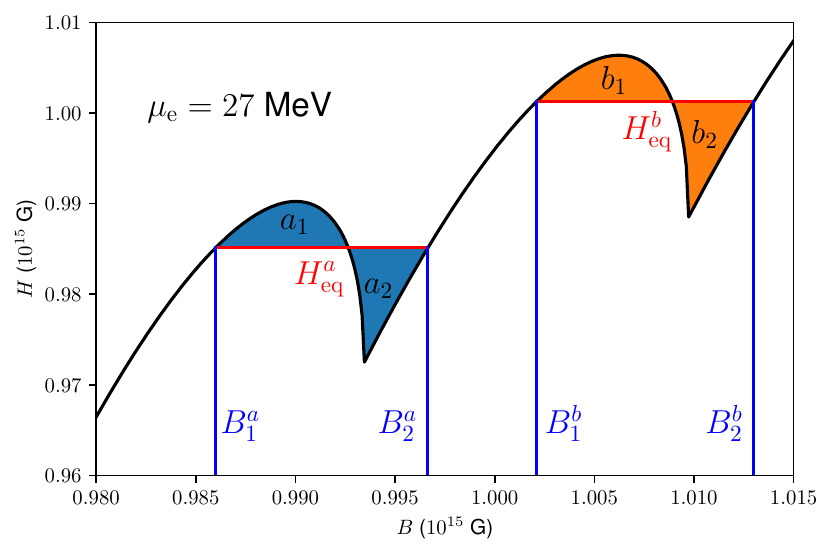}
\caption[]{$H(B)$ for fixed $\mu_{\text{e}}$ and $T=0$, showing how the Maxwell construction is used to determine the equilibrium value of the field $H$ and the span of $B$ for which $H$ is uniform. The construction has been applied to two different domain-forming regions of parameter space distinguished by superscripts $a$ and $b$. Shaded areas $a_1$ and $a_2$, and areas $b_1$ and $b_2$, are equal.}
\label{fig:MaxwellConstructionH}
\end{figure}

\begin{figure}
\centering
\subfloat[]{\includegraphics[width=0.5\linewidth]{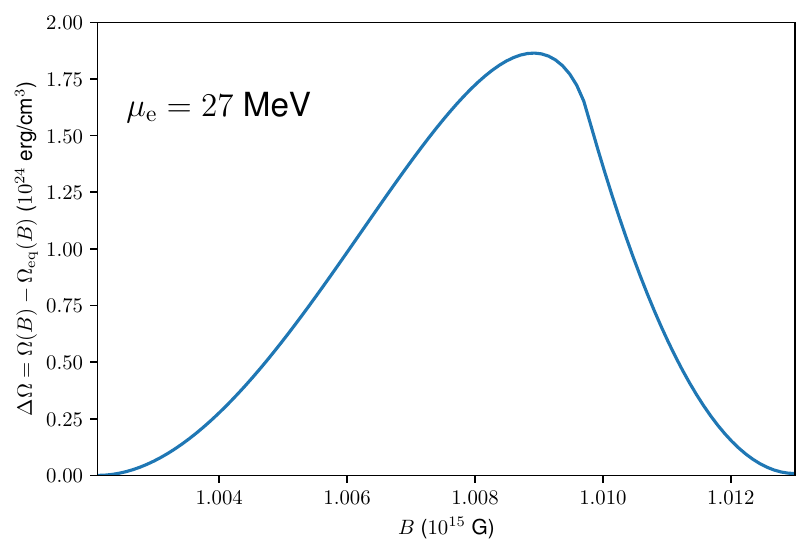}}
  \hfill
  \subfloat[]{\includegraphics[width=0.489\linewidth]{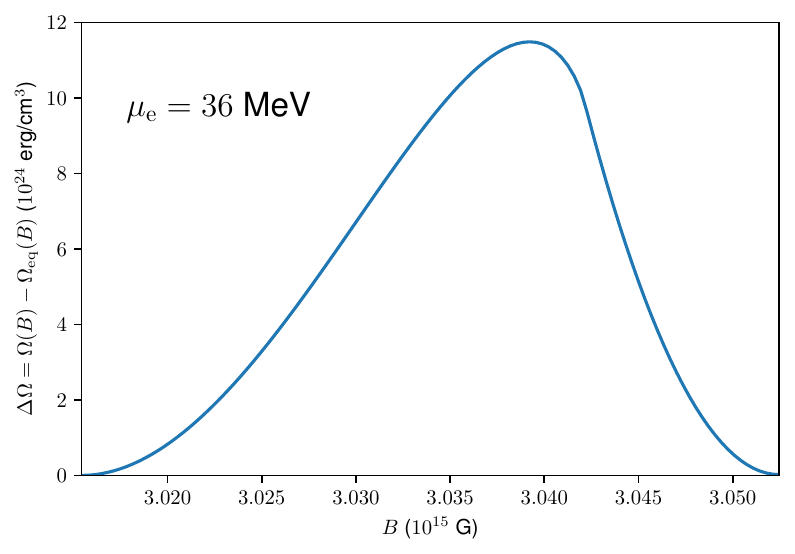}}
\caption[]{The difference in the grand potential density between the unstable (no domains) and stable equilibrium (with domains) state at two distinct domain-forming ranges of $B$ for fixed $\mu_{\text{e}}$ and $T=0$. Note the difference fixed values of $\mu_{\text{e}}$ for panels (a) and (b).}
\label{fig:Deltau}
\end{figure}

In the unstable regions, the equilibrium configuration is found using the Maxwell construction for $H(B)$ at fixed $\mu_{\text{e}}$: $\mu_{\text{e}}$ must be held constant so that electrons can be exchanged between neighbouring domains without energy cost. Working at $T=0$ for simplicity, the equilibrium value of $H$ in the unstable region is found by finding the value of $H$ at which the thermodynamic potential
\begin{equation}
\overline{\Omega}=\Omega-\frac{H^iB_i}{4\pi},
\end{equation}
intersects itself. If the equilibrium value of $H$ in a domain-forming region is $H_{\text{eq}}$, and the two values of $B$ such that $H(B)=H_{\text{eq}}$ and $(\partial H/\partial B)_{n_{\text{e}}}>0$ bounding the domain-forming region are $B_1$ and $B_2$ where $B_1<B_2$, then the grand potential density in the domain-forming region of parameter space at a stable equilibrium is
\begin{equation}
\Omega_{\text{eq}}=\Omega(B_1)+\frac{1}{4\pi}H_{\text{eq}}(B-B_1).
\end{equation}
$\Omega_{\text{eq}}$ joins continuously to $\Omega$ outside this range of $B$. The slope $(\partial \Omega/\partial B)_{\mu_{\text{e}}}=H/(4\pi)$ is also continuous, but $(\partial H/\partial B)_{\mu_{\text{e}}}$ is discontinuous. The Maxwell construction applied to $H(B)$ is shown in Figure~\ref{fig:MaxwellConstructionH}, and the energy difference $\Delta \Omega=\Omega-\Omega_{\text{eq}}$ between the unstable and equilibrium configuration in a domain-forming region is shown in Figure~\ref{fig:Deltau}. Note that the sections of the original $H(B)$ curve included within the domain-forming regions where $1-4\pi\chi_n>0$ or equivalently $(\partial H/\partial B)_{n_{\text{e}}}>0$ are metastable, and domains do not necessarily form in this region of parameter space absent nucleation sites for domain walls.

~\citet{Blandford1982} discuss how the electrodynamic boundary conditions at the no-domain to domain-forming region interface can modify the picture discussed above. In the absence of surface currents, the electrodynamic boundary conditions are
\begin{equation}
\hat{n}_iB^i_1=\hat{n}_iB^i_2,\quad \epsilon_{ijk}\hat{n}^jH_1^k=\epsilon_{ijk}\hat{n}^jH_2^k,
\end{equation}
for $\hat{n}^i$ normal to the no-domain to domain-forming region interface and $1$ and $2$ denote these distinct regions. Under the assumption that the angle $\theta$ between $B$ and the interface varies only by a small amount $\delta\theta\ll\theta$ between the two sides of the interface and that the magnitudes of $H^i$ and $B^i$ are similar on each side, one can show that
\begin{equation}
\frac{H_2-H_1}{B_2-B_1}\approx -\tan^2\theta,
\label{eq:HBBoundaryConditions}
\end{equation} 
and hence the Maxwell construction as used above is only correct if $\theta=0$ and $H_1=H_2$. The correct Maxwell construction in the general case is
\begin{equation}
\int^{B_2}_{B_1}H(B)\textrm{d}B=\frac{1}{2}(H_1+H_2)(B_2-B_1),
\label{eq:MaxwellConstruction}
\end{equation}
where $H_1=H(B_1)$ and $H_2=H(B_2)$ are the values of $H$ at the two ends of the linear ``equilibrium'' solution for $H(B)$. $H_2\leq H_1$ by Eq.~(\ref{eq:HBBoundaryConditions}), so if $H_1\neq H_2$, then the two points on the $H(B)$ curve which are connected by the ``equilibrium'' solution are metastable, and the ``equilibrium'' solution is still thermodynamically unstable since $(\partial H/\partial B)_{n_{\text{e}}}<0$ in the range $B_1<B<B_2$. The case where $H_1=H_2$ is energetically favourable and is the true equilibrium unless additional forces not considered here (e.g., force balance between the Lorentz force and crustal elasticity) prevent it from being realized.

Unlike the previous studies on neutron star crusts, we are less interested in the magnetostriction associated with the magnetic domains and its ability to e.g., ``crack'' the crust, than we are with a potential mechanism to release heat during the domain formation process. However, a newly-born magnetar will not initially contain magnetic domains due to their suppression at high temperatures. As the star cools below $T_{\text{crit}}$ (which depends on $B$ and $n_{\text{e}}$), domain formation is possible if $B$ and $n_{\text{e}}$ in a particular region of the star are such that a Landau level is nearly empty/almost filled. The timescale to form these domains is of order the Ohmic decay timescale~\citep{Blandford1982} as the domain formation process is dominated by eddy currents.

The instability of both the Hall modes and the Alfv\'{e}n modes as discussed in Section~\ref{sec:HallMHD} requires $\chi_n>1/(4\pi)$ to be true, so it is only possible in a region of parameter space which is already thermodynamically unstable. Since the timescale of this instability is shorter than the Ohmic decay timescale, which is also the domain formation timescale, this suggests that unstable field growth through this MHD instability could occur before domain formation locally stabilizes the system. Since the growth time of the unstable Alfv\'{e}n modes is much shorter than for the unstable Hall modes, the unstable field growth would be dominated by the former. Because this field growth would occur over length scales of order the domain size, which is of order the spacing between consecutive regions where new Landau levels become occupied $\sim10-100$ cm, it could lead to enhanced Ohmic dissipation, since these length scales are relatively short in the context of neutron star crusts. Additionally, metastable phases of non-uniform magnetization can exist in the absence of domain wall nucleation site. If disturbed, domain formation and a release of energy is possible in these regions. However, we emphasize that the macroscopic fluid equations that are the basis of this paper are insufficient to explain domain formation, which requires a more microphysical treatment.

We consider the possibility of the dynamic destabilization of the crust due to thermal or magnetic field evolution, and the resulting heating by Ohmic dissipation of magnetic fields that grow unstably due to the Hall MHD instability described in Section~\ref{sec:HallMHD}. We can make a ``back-of-the-envelope'' estimate of the maximum heating that the domain formation process could provide. The maximum energy per unit volume available to be released during the domain formation process is $\Delta\Omega$. Not all of this will be available to be converted into field growth and then decay to heat the crust, as forming domain walls requires energy, though this will be of order a percent of the total energy available~\citep{Wang2016}. 

The heating occurs in the unstable regions which are spaced throughout the crust, which we approximate as spherical shells. Their radial extent $\Delta z$ varies from order tens to hundreds of cm, becoming more compressed where the field is weaker at fixed density or where the density is increased at fixed field. To determine how many of these regions there is, we assume a uniform field strength $B_0$ and note that the crust extends from $\mu_{\text{e}}=m_{\text{e}}$ to $\mu_{\text{e},\text{cc}}\approx70$ MeV, so starting from the surface and moving inward successive Landau levels from $n=0$ to
\begin{equation}
n_{\text{max},\text{cc}}=\left\lfloor\frac{\mu_{\text{e},\text{cc}}^2-m_{\text{e}}^2}{2eB}\right\rfloor=\left\lfloor\frac{\mu_{\text{e},\text{cc}}^2-m_{\text{e}}^2}{11.83B_{15}\text{ MeV}^2}\right\rfloor,
\end{equation}
will be occupied. For $B_{15}=B/(10^{15}\text{ G})=1$, $n_{\text{max},\text{cc}}\approx400$, so there are 400 regions which are potentially unstable to domain formation. We know from Section~\ref{sec:NonzeroTUnstablePS} that nonzero temperature will suppress the instability in many of these regions, so we use a reduced value $N_{\text{in}}\approx 100$. We also make the very optimistic assumption that these regions are all destabilized simultaneously \textit{and} that all have similar values of $\Delta \Omega$.

The relevant timescale for the heating through this mechanism is the Ohmic dissipation timescale
\begin{equation}
\tau_{\text{O}}=\frac{4\pi L^2\sigma}{c^2},
\end{equation}
for length scale $L$ and conductivity $\sigma$, not the much shorter instability timescale. We take $L\approx\Delta z$. The conductivity of a neutron star crust ranges from $\sigma\approx 10^{17}$--$10^{27}$ s$^{-1}$~\citep{Potekhin1999}; for temperatures $\sim 5\times10^8$ K, in the bulk of the crust $\sigma\sim 10^{23}$ s$^{-1}$. The heat flux from this mechanism is thus
\begin{equation}
F_{\text{O}}\approx N_{\text{in}}\frac{\Delta\Omega\Delta z}{\tau_{\text{O}}}\approx 3.3\times10^{21}\left(\frac{N_{\text{in}}}{100}\right)\left(\frac{10^{24}\text{ s}}{\sigma}\right)\left(\frac{\Delta \Omega}{10^{23}\text{ erg/cm}^3}\right)\left(\frac{100\text{ cm}}{\Delta z}\right)\frac{\text{erg}}{\text{cm}^2\text{ s}},
\end{equation}
assuming that the heating is continuous, though this is not true since it ceases once the domains form.~\citet{Beloborodov2016} estimated that for magnetars to sustain their strong surface flux, the heating mechanism in their crust must provide a heat flux of order $F\sim 10^{24}(\epsilon/0.01)^{-1}$ erg cm$^{-2}$ s$^{-1}$, where $\epsilon$ is an efficiency factor accounting for heat loss to the crust and to neutrino emission. Thus, even based on our very optimistic analysis, magnetic domain-associated Ohmic dissipation is far too small to be of any significance as a crustal heating mechanism.

\subsection{Location of domains}
\label{sec:DomainLocation}

In the ultrarelativistic limit of massless electrons $p_F\approx\mu_{\text{e}}$, which is an excellent approximation for most of a neutron star crust, we can determine the location in $B$--$\mu_{\text{e}}$ parameter space of the magnetic domain-forming regions relatively easily. Under this approximation we can write $P_{\text{e}}$ and $M$ as
\begin{align}
P_{\text{e}}={}&\frac{eBp_F^2}{2\pi^2}\sum_{n=0}^{n_{\text{max}}}\frac{g_n}{2}\left[\sqrt{1-bn}-bn\, \text{arsinh}\left(\sqrt{\frac{1}{bn}-1}\right)\right]=\frac{p_F^4}{4\pi^2}b\left[\Sigma_{1/2}(b)-\Sigma_{S,n}(b)\right],
\label{eq:PeApprox}
\\
M={}&\frac{ep_F^2}{2\pi^2}\sum_{n=0}^{n_{\text{max}}}\frac{g_n}{2}\left[\sqrt{1-bn}-2bn\, \text{arsinh}\left(\sqrt{\frac{1}{bn}-1}\right)\right]=\frac{ep_F^2}{2\pi^2}\left[\Sigma_{1/2}(b)-2\Sigma_{S,n}(b)\right],
\label{eq:MApprox2}
\end{align}
where $b\equiv 2eB/p_F^2$, and where $\Sigma_{1/2}(b)$ and $\Sigma_{S,n}(b)$ are given by $\Sigma_{1/2}$ and $\Sigma_{S,n}$ in Eq.~(\ref{eq:SigmaExpressions}) with $n_{\text{max}}\rightarrow b^{-1}$ and including the $n=n_{\text{max}}$ term in the sum. Note that these equations are functions of only a single parameter $b$. For fixed $\mu_{\text{e}}$, Eq.~(\ref{eq:MaxwellConstruction}) implies that
\begin{equation}
\frac{P_{\text{e}}(B_2,\mu_{\text{e}})-P_{\text{e}}(B_1,\mu_{\text{e}})}{B_2-B_1}=\frac{1}{2}\left(M(B_1,\mu_{\text{e}})+M(B_2,\mu_{\text{e}})\right).
\label{eq:DomainCondition}
\end{equation}
Note that this is inconsistent with Eq.~(17) of~\citet{Blandford1982}: we find that their Eq.~(17) is only possible if the two possible magnetizations in the domain-forming regions are of equal magnitude but opposite direction, but this is not true in general.

\begin{figure}
\centering
\includegraphics[width=0.5\linewidth]{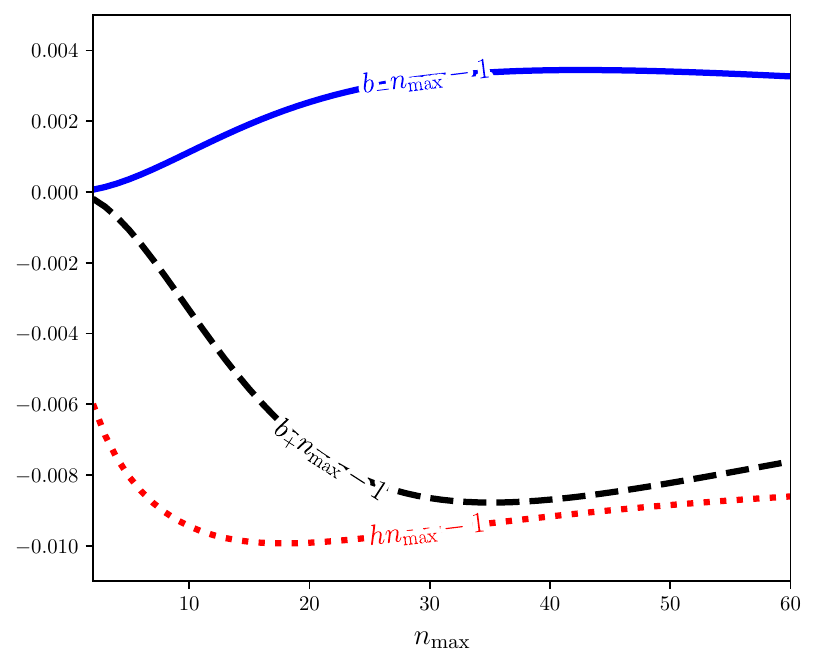}
\caption[]{Scaled, shifted values of $b=2eB/p_F^2\approx 2eB/\mu_{\text{e}}^2$ as a function of $n_{\text{max}}$ at the boundaries of the domain-forming regions, with subscript $+$ and $-$ denoting the upper and lower boundary values respectively. The constant values of $h=2eH/p_F^2$ in the domain-forming regions, scaled and shifted by the same factors as the $b$ values, are also displayed.}
\label{fig:DomainLocations}
\end{figure}

Using Eq.~(\ref{eq:PeApprox}--\ref{eq:MApprox2}), the locations of the domains as a function of $b$ can be readily obtained. The upper and lower bounding values of $b$ of the domain-forming regions are plotted as a function $n_{\text{max}}$ in Figure~\ref{fig:DomainLocations}. The constant values of $h=2eH/p_F^2$ in the domain-forming regions is also shown. Since $b$ is a ratio between $B$ and $p_F^2\approx\mu_{\text{e}}^2$, these values only need to be computed once; then for a given $B$, they can be translated into the ranges of $\mu_{\text{e}}$ where domain formation will occur or \textit{vice versa}.

\section{Discussion and Conclusion}
\label{sec:Conclusion}

Magnetohydrodynamic instabilities within magnetars are commonly invoked in mechanisms for energetic magnetar outbursts, including many proposed fast radio burst models. These instabilities could cause crust yielding and hence magnetospheric twisting and bursting activity. The thermal and magnetic evolution of magnetars are intricately linked, and the decay of strong surface fields typical of magnetars possibly powers a heating mechanism responsible for the higher luminosities observed for magnetars compared to rotation-powered pulsars. We thus studied Hall MHD applied to a neutron star crust threaded by magnetic fields in the range characteristic of magnetars, with the goal of looking for instabilities that could affect magnetic field evolution or crustal heating.

Like in the core, we find instabilities associated with the population of new Landau levels originating from the differential magnetic susceptibility. These are hydrodynamic instabilities with thermodynamic origins, and only occur in regions of the star unstable to magnetic domain formation. After domains form, the instability is absent. However, when this happens the wave vector of a mode of a particular frequency will change across a domain boundary due to discontinuous $M$ and $\chi_n$, and hence refraction of MHD modes will occur at the domain walls. In the crust, the novel instability is associated with the circularly-polarized Hall modes, though like in the core we also find unstable Alfv\'{e}n modes, which grow much faster. Using linear perturbation theory we estimated the Hall mode instability's growth time, finding a slow-growing mode with timescale set approximately by the inverse Hall mode frequency. In these same regions, unstable Alfv\'{e}n modes grow on timescales of order the local Alfv\'{e}n mode period. We showed that the instability is self-limiting; unstable field growth will eventually stabilize the magnetized fluid locally and magnetic domain formation will occur. It is also suppressed above a field-and density-dependent critical temperature. This suggests the instability will only become active after the magnetar cools sufficiently, to temperatures $\sim10^7$--$10^9$ K depending on the location within the crust and strength of the magnetic field. In particular, we showed (see Figure~\ref{fig:dHdBandTcritvsnmax} panel (b)) that the peak value of the critical temperature for instability is about $0.0015p_{F}$ where electrons are relativistic, and it happens at $\nmax=p_F^2/2eB=45$. This also means that at a given $B$ the peak temperature for stability is where $p_F\approx\sqrt{90eB}$ and therefore has a value of about $0.0015\sqrt{90eB}=8.4\times10^7\sqrt{B/B_q}$ K where $B_q=m_{\text{e}}^2/e$.

We studied the possibility that the domain formation transition could be a heat source in magnetar crusts. We estimated the Ohmic dissipation of magnetic fields that grow by the Hall MHD instability described in this paper in unstable regions of the crust. Even using an optimistic estimate, this mechanism is far too intermittent, and the energy available to be converted into magnetic field and then heat via Ohmic dissipation is many orders of magnitude too small, to dominate magnetar crustal heating. We also showed that the magnetization alone can increase the Ohmic dissipation by a factor of a few in a strongly magnetized neutron star due to the oscillations of the differential magnetic susceptibility. This effect is independent of the instability and persists at temperatures for which the instability is suppressed, and also to regions of field-density parameter space where the instability is not active (i.e., the strongly-quantized regime). However, this enhancement will only occur in limited spatial regions, so its overall effect on magneto-thermal evolution of neutron stars may be modest. Numerical simulations may be necessary to determine whether this effect can have observable implications. An additional possible heating source is viscosity in the plastic flow, though the uncertain strength of the viscosity as discussed in Section~\ref{sec:PlasticFlow} means that further study is necessary to determine its overall significance.

\section{Acknowledgements}
We thank the anonymous referee for useful comments. P.~B.~R. was supported by the INT's U.S. Department of Energy grant No. DE-FG02-00ER41132.

\section*{Data Availability}

There are no new data associated with this article.

\bibliography{library,textbooks,librarySpecial}

\bsp	
\label{lastpage}
\end{document}